\documentclass[prl,twocolumn,superscriptaddress,nofootinbib]{revtex4-1}

\usepackage{graphicx,color,amsmath}
\usepackage[export]{adjustbox}
\usepackage{slashed}
 \usepackage{url}
\newcommand{\eV}{{\, {\rm eV}}}

\usepackage{tikz}
\usepackage{tkz-euclide}
\usetkzobj{all}
\usetikzlibrary{decorations.pathmorphing}	% To create gluons and vector bosons
\tikzset{
    v/.style={decorate, decoration={snake, segment length=3mm, amplitude=0.75mm}, draw},
    f/.style={draw=black, postaction={decorate},
        decoration={markings,mark=at position .6 with {\arrow[very thick]{latex}}}},
    fb/.style={draw=black, postaction={decorate},
        decoration={markings,mark=at position .4 with {\arrowreversed[very thick]{latex}}}},
    fnar/.style={draw=black},
    g/.style={decorate, draw=black,
        decoration={coil,amplitude=3pt, segment length=3.5pt}},
    s/.style={dashed,draw=black, postaction={decorate},
        decoration={markings,mark=at position .55 with {\arrow[very thick]{latex}}}},
    sb/.style={dashed,draw=black, postaction={decorate},
        decoration={markings,mark=at position .55 with {\arrowreversed[draw=black,very thick]{latex}}}},
    snar/.style={dashed,draw=black,line width =1.25pt},
}

\definecolor{mypurple}{RGB}{164,64,214}

\newcounter{qnumber}

\begin{document}

\title{New Physics in the Rayleigh-Jeans Tail of the CMB}

\affiliation{Perimeter Institute for Theoretical Physics, 31 Caroline Street N, Waterloo, Ontario N2L 2Y5, Canada}
\author{Maxim Pospelov}
%\email{mpospelov@perimeterinstitute.ca}
\affiliation{Perimeter Institute for Theoretical Physics, 31 Caroline Street N, Waterloo, Ontario N2L 2Y5, Canada}
\affiliation{Department of Physics and Astronomy, University of Victoria, Victoria, BC V8P 5C2, Canada} 
\affiliation{Theoretical Physics Department, CERN, Geneva, Switzerland}
\author{Josef  Pradler}
\affiliation{Institute of High Energy Physics, Austrian Academy of Sciences, Nikolsdorfergasse 18, 1050 Vienna, Austria}
\author{Joshua T. Ruderman}
%\email{mpospelov@perimeterinstitute.ca}
\affiliation{Center for Cosmology and Particle Physics, Department of Physics,
New York University, New York, NY 10003, USA}
\affiliation{Theoretical Physics Department, CERN, Geneva, Switzerland}
\author{Alfredo Urbano}
%\email{mpospelov@perimeterinstitute.ca}
\affiliation{INFN, sezione di Trieste, SISSA, via Bonomea 265, 34136 Trieste, Italy}
\affiliation{Theoretical Physics Department, CERN, Geneva, Switzerland}

\date{\today}

\begin{abstract}

  We show that despite stringent constraints on the shape of the main
  part of the CMB spectrum, there is considerable room for its
  modification within its Rayleigh-Jeans (RJ) end,
  $\omega \ll T_{\rm CMB}$.  We construct explicit New Physics models
  that give an order one (or larger) increase of
  photon count in the RJ tail, which can be tested by existing and
  upcoming experiments aiming to detect the cosmological 21\,cm
  emission/absorption signal. This class of models stipulates the
  decay of unstable particles to dark photons, $A'$, that have a small
  mass, $m_{A'} \sim 10^{-14} - 10^{-9}$\,eV, non-vanishing mixing
  angle $\epsilon$ with electromagnetism, and energies much smaller
  than $T_{\rm CMB}$. The non-thermal number density of dark photons
  can be many orders of magnitude above the number density of CMB
  photons, and even a small probability of $A'\to A$ oscillations,
  for values of 
  as small as $\epsilon \sim 10^{-9}$, can significantly increase the
  number of RJ photons. In particular, we show that resonant
  oscillations of dark photons into regular photons in the interval of
  redshifts $20 < z < 1700$ can be invoked as an explanation of the
  recent tentative observation of a stronger-than-expected absorption
  signal of 21\,cm photons.  We present a model that realizes this
  possibility, where milli-eV mass dark matter decays to dark photons,
  with a lifetime longer than the age of the Universe.
\end{abstract}

\maketitle

%%%%%%%%%%%%%%%%%%%%%%%%%%%%%%%%%%%%%%%%%%%%%%%%%%

\paragraph*{ Introduction:}

Modern cosmology owes much of its advance to precision observations of
the Cosmic Microwave Background (CMB)\@. Both the spectrum of the CMB
and its angular anisotropies are precisely measured by a number of
landmark experiments \cite{Mather:1993ij,Hinshaw:2012aka,Ade:2015xua}.
CMB physics continues its advance \cite{Abazajian:2016yjj} into
probing both the standard $\Lambda$CDM model to higher precision and
possible New Physics that can manifest itself as small deviations from
theoretical expectations. In addition, a qualitatively new
cosmological probe, the physics of 21\,cm emission/absorption at the
end of the ``dark ages," may come into play in the very near future~\cite{Furlanetto:2006jb}.

Cosmology has been a vital tool for learning about physics beyond the
Standard Model (SM)\@.  About one quarter of our Universe's energy
budget is comprised of cold Dark Matter (DM).
The precision
tools of cosmology provide serious constraints on the properties of
DM, which instead of coming ``alone," may be a part of an extended
dark sector, comprising new matter and radiation fields, and new
forces. Recent years have seen a significant increase in studies of
dark sectors, both in connection with terrestrial experiments, and in
cosmology~\cite{Jaeckel:2010ni,Alexander:2016aln,Battaglieri:2017aum}.

If such light fields are thermally excited, they can be detected
through their gravitational interaction alone, as they would modify
the Hubble expansion rate, affect the outcome of Big Bang
Nucleosynthesis, 
and modify the angular anisotropies of the CMB\@. The resulting
constraint
is phrased in terms of the number of effective neutrino degrees of
freedom which, according to the latest observational bounds,
$N_{\rm eff} = 3.04 \pm 0.33$~\cite{Ade:2015xua} (95\% C.L.), is consistent with the
expectations 
from standard cosmology.  Non-thermal Dark Radiation
(DR) is considered in the literature less often, although many dark
sector processes may lead to its appearance.

\begin{figure}[!t!]
\centering
\includegraphics[width=0.5\textwidth]{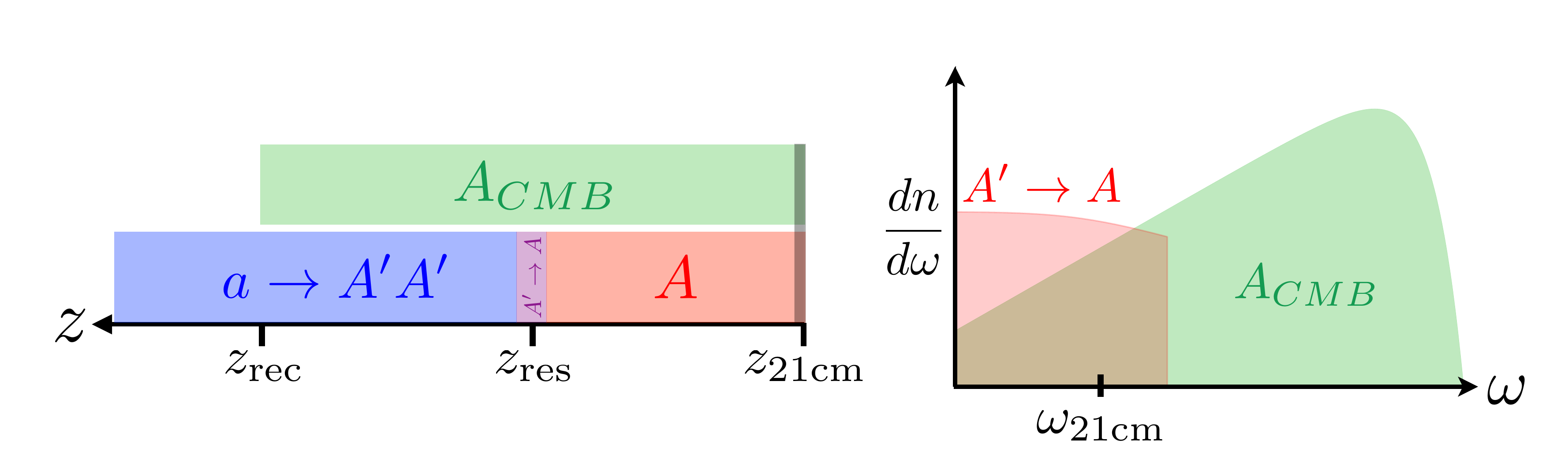}  
\caption{\label{fig:Schematic} Dark matter, $a$, decays to soft
  massive dark photons, $A'$, which oscillate into ordinary photons,
  $A$.  The conversion happens resonantly at
  redshift $z_{\rm res}$
  before the formation of the 21\,cm absorption signal.  The converted
  photons (red) add to the CMB photon count (green) in the RJ tail.}
\end{figure}

\begin{figure*}[!ht!]
\centering
\includegraphics[width=.42\textwidth]{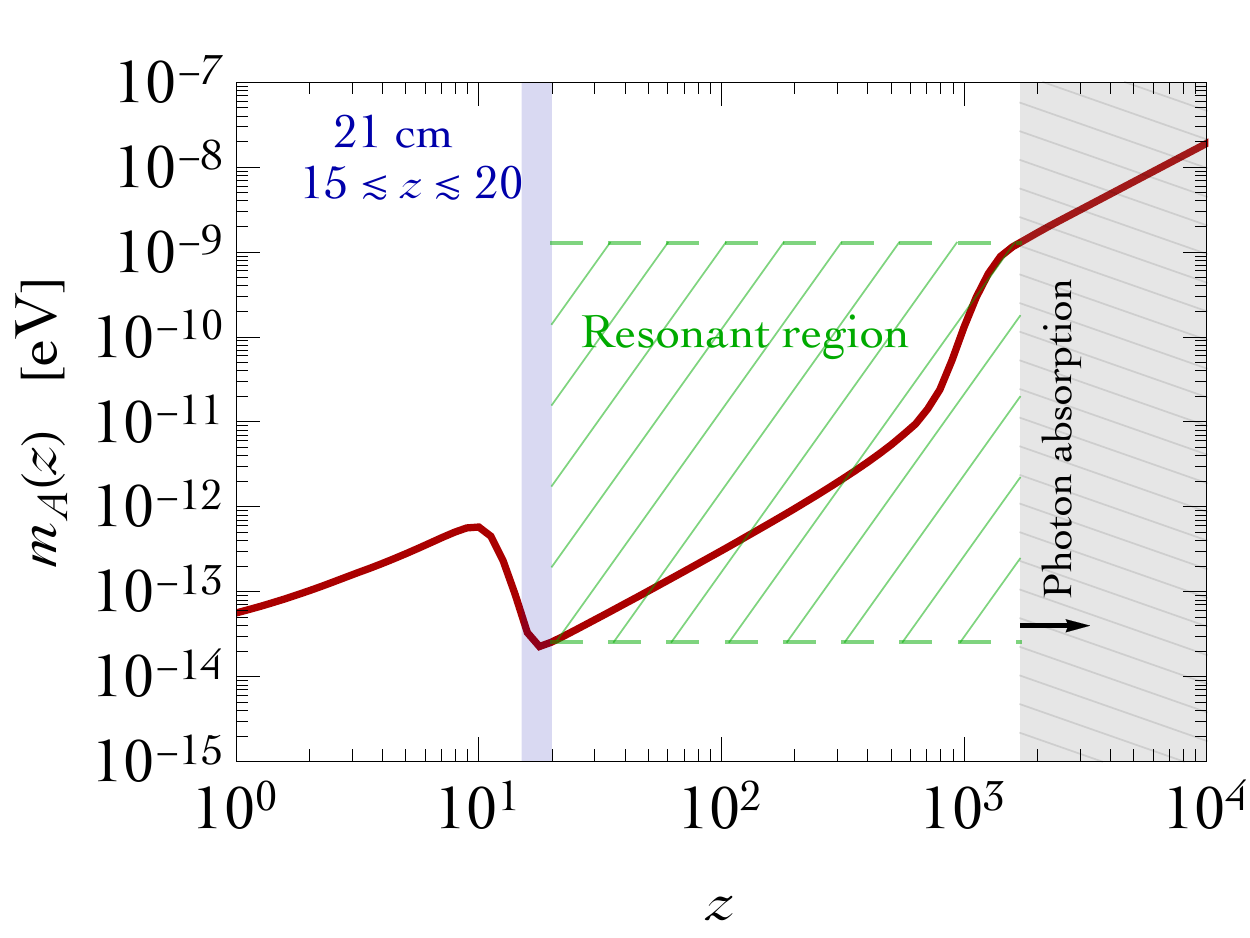}%
\hspace*{0.5cm}
  \includegraphics[width=.42\textwidth]{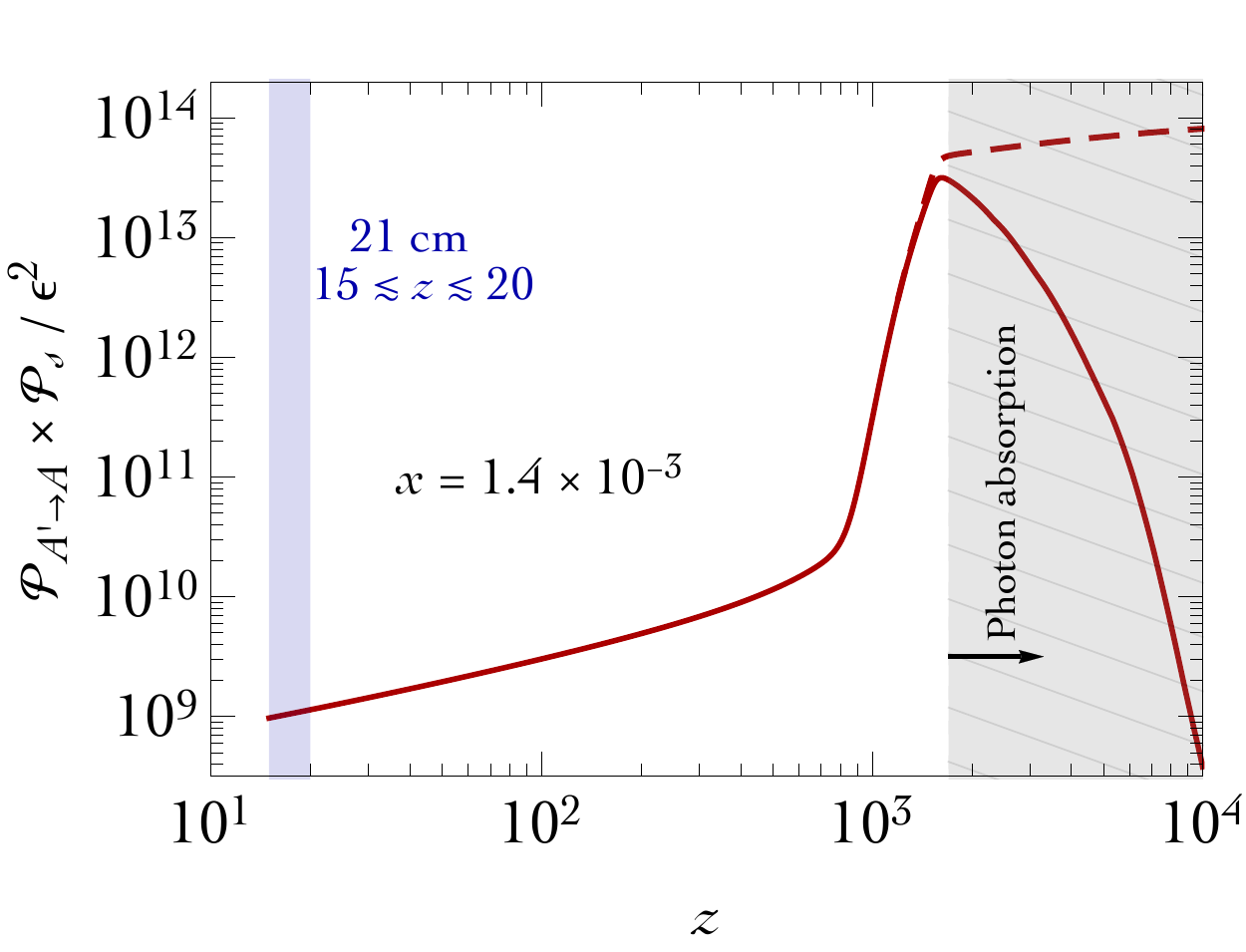}%
\vspace{-.2 cm}
\caption{\label{fig:MassAndProb}
  {\em Left.}  Effective photon mass $m_A$ as a function of redshift.
  When resonant oscillation [$m_A(z)=m_{A'}$] occurs between redshift $20 \leq z \lesssim 1700$ (green) it impacts the 21\,cm
  absorption signal.
  {\em Right.} Effective conversion probability as a function of
  redshift.  We consider photons with energy
  $x = \omega / T_{CMB} = 1.4 \times 10^{-3}$, which implies a
  wavelength of 21\,cm at $z = 17$. }
\end{figure*}

In recent papers~\cite{Cherry:2015oca,Cui:2017ytb}, {\em interacting }
DR was examined in the regime where the individual quanta are fewer in
number but harder in 
energy than typical CMB photons,
$ n_{\rm DR} \ll n_{\rm CMB}$ and $\omega_{\rm DR} \gg \omega_{\rm CMB}$,
but such that the $N_{\rm eff}$ constraint is satisfied.
In
this letter we study the alternative,
DR quanta much softer, but more numerous than CMB photons,
\begin{equation}
   \omega_{\rm DR} \ll \omega_{\rm CMB},\quad  n_{\rm DR} > n_{\rm RJ}  , \quad \omega_{\rm DR}n_{\rm DR} \ll \rho_{\rm tot},
\label{setup}
\end{equation}
where $\rho_{\rm tot}$ is the total energy density of radiation and DM, 
$n_{\rm DR}$ is the number density of DR quanta, and $n_{\rm RJ}$ represents the 
low-energy Rayleigh-Jeans (RJ) tail of the standard CMB Planck distribution, $n_{RJ} \approx T\omega^2_{\rm max} / (2 \pi^2) \approx 0.21 x_{\rm max}^2n_{\rm CMB} $,
where we use units 
$\hbar=c=k=1$, and where we define the normalized 
 photon energy, $x \equiv \omega / T_{\rm CMB}$.
$n_{\rm CMB} = 2\zeta(3)/\pi^2\,T_{\rm CMB}^3 \simeq 0.24\,T_{\rm
  CMB}^3$ is the full Planckian number density and
$x_{\rm max} = \omega_{\rm max} / T_{\rm CMB}$ is a (somewhat
arbitrary) maximum frequency of the low-energy RJ interval,
$x_{\rm max} \ll 1$.  If for example we take
$x_{\rm max} =2\times 10^{-3}$, then we find
$n_{\rm RJ}/n_{\rm CMB}\simeq 10^{-6}$.

The number of DR quanta may significantly exceed $n_{\rm
  RJ}$. Saturating the constraint on $N_{\rm eff}$ for the DR that matches
the CMB frequencies with $x_{\rm max} \sim 2\times 10^{-3}$, we find
$n_{\rm DR} \leq 1.5 \times 10^2 \, n_{\rm CMB}$.
This is because the energy density of a DR
 component can be written as
$\rho_{\rm DR} = \omega_{\rm DR}n_{\rm DR} = \Delta N_{\rm eff}\frac{7}{8}(\frac{4}{11})^{4/3}\rho_{\rm CMB}$, where we separated the SM contribution $N_{\rm eff} = N^{\rm SM}_{\rm eff} + \Delta N_{\rm eff}$
with $N^{\rm SM}_{\rm eff} =3.046$.
  Alternatively,
letting $5\%$ of the DM energy density
\cite{Berezhiani:2015yta,Poulin:2016nat} convert to DR in the
same frequency range after  CMB decoupling,
$n_{\rm DR} \leq 3.3 \times 10^{5} \, n_{\rm CMB}$.
Thus, soft DR quanta can outnumber the RJ CMB photons by up to 11 orders of magnitude. 

\begin{figure*}[!t!] 
\centering
 \raisebox{0.075cm}{\includegraphics[width=.38\textwidth]{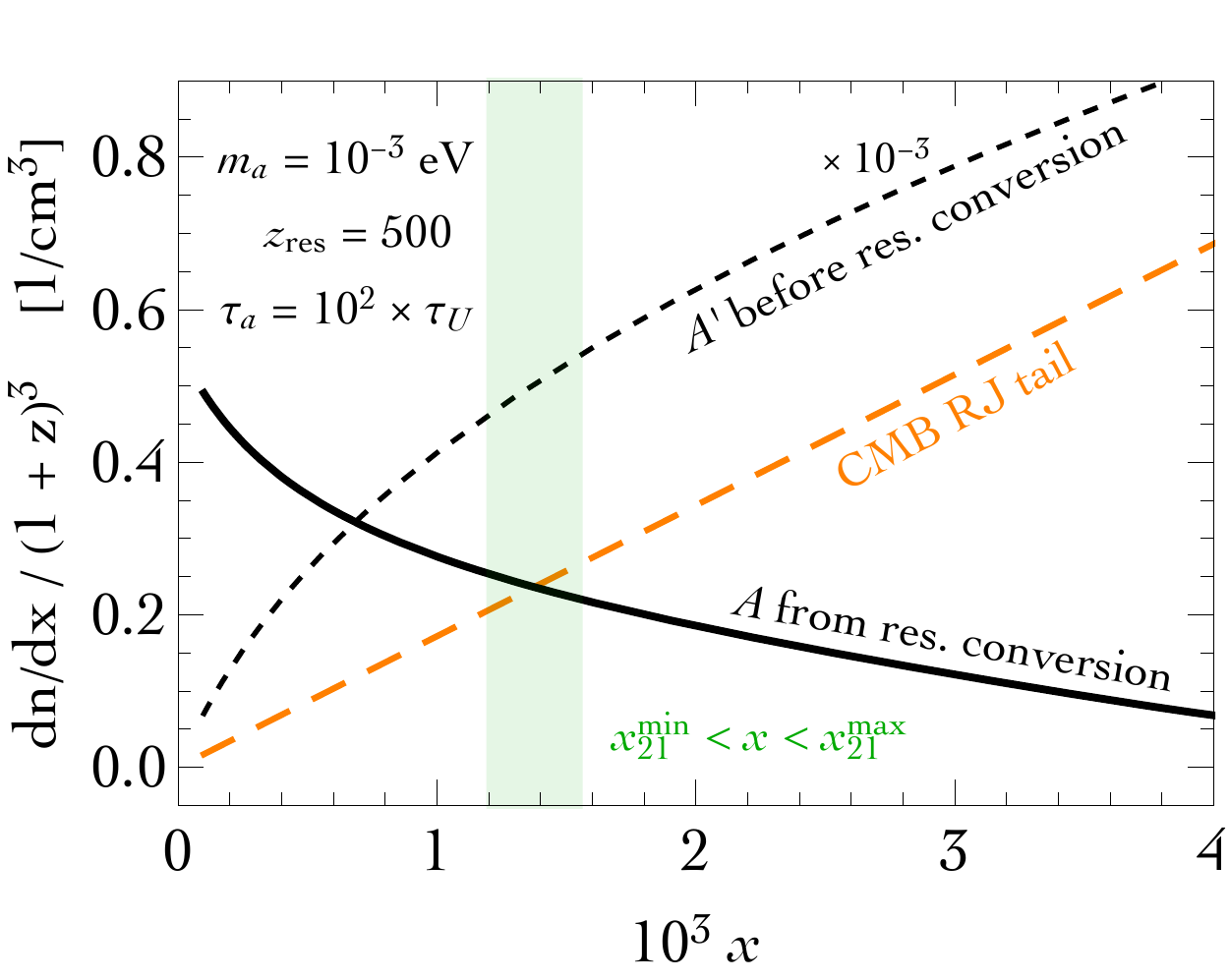}}
\hspace{0.5cm}
\includegraphics[width=.41\textwidth]{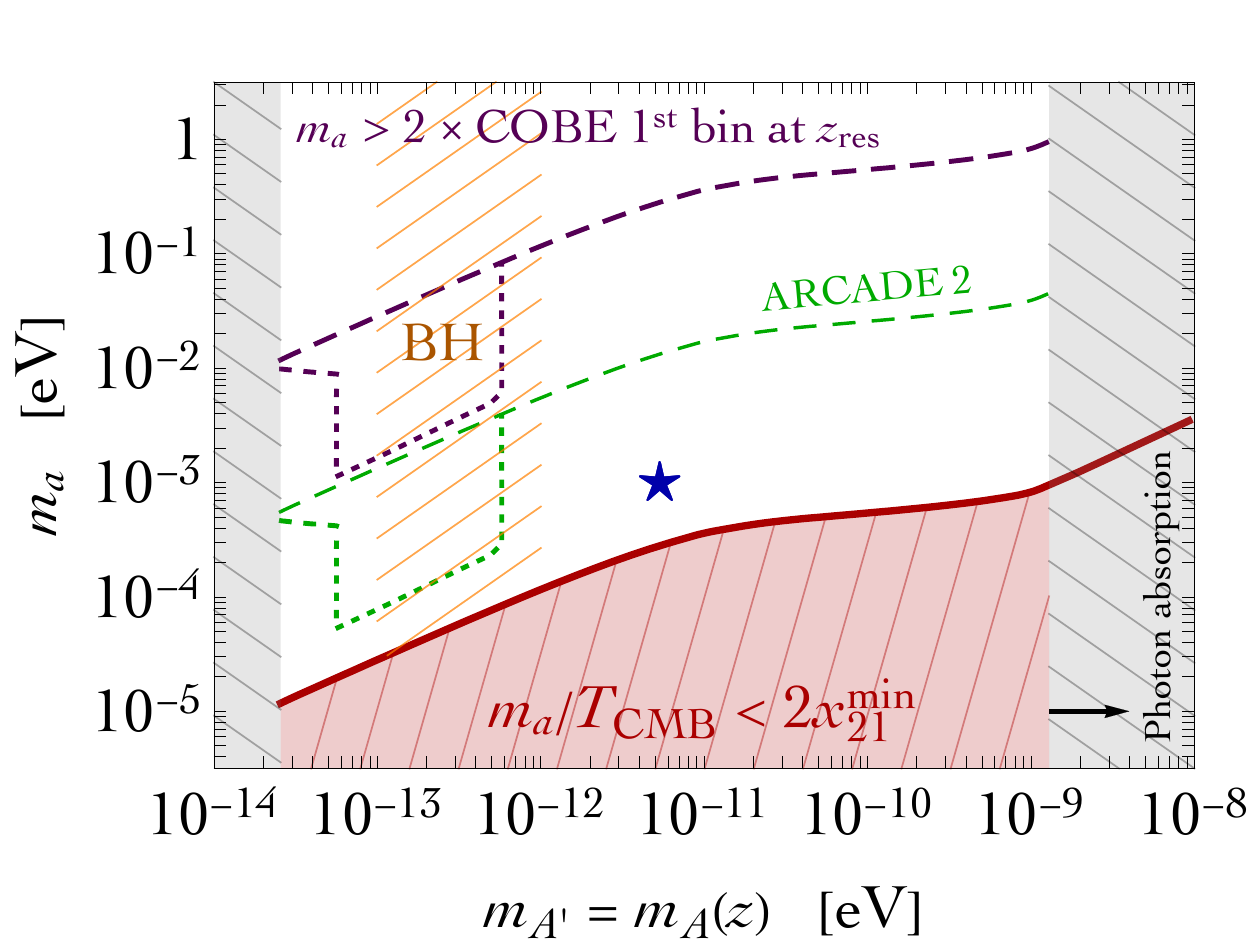}
\vspace{-.2 cm}
\caption{ \label{fig:BoundMassAndSpectrumAfterConversion}
  {\em Left.} (Dark) photon spectrum (before) after conversion shown
  in solid black (dashed black, rescaled by $10^{-3}$) and compared to
  the CMB (orange) for $m_a = 10^{-3}\,\mathrm{eV}$,
  $\tau_{a} = 100\times \tau_{\rm U}$,
  $m_{A'} = 5 \times 10^{-12}~\mathrm{eV}$---implying resonance at
  $z_{\rm res} = 500$---and $\epsilon = 2.1 \times 10^{-7}$.  {\em
    Right.}  The relevant range of DM mass, $m_a$, versus $m_{A'}$ for addressing the
  21\,cm absorption signal; for $m_a/T_{\rm CMB}< 2x_{21}^{\rm min}$
  the produced photons are too soft to impact 21\,cm.  The region
  above the dashed purple (green) line are probed by FIRAS (ARCADE 2),
  constraining $\epsilon$ and $\tau_a$.  The BH region is possibly
  disfavored by black hole superradiance~\cite{Cardoso:2018tly}.  The
  star shows the example point considered on the left panel.}
\end{figure*}

What are the observational consequences of such soft and numerous DR
quanta?  Very light fields often have their interactions enhanced
(suppressed) at high (low) energies. This is the case for neutrinos,
that have Fermi-type interactions with atomic constituents, as well as
for
axions that have effective interactions with fermions and gauge
bosons.  This type of DR would be very difficult to see directly.
There is, however, one class of new fields that can manifest their
interactions at low energies and low densities.  These are light
vector particles (often called dark photons), $A'$, that develop
a mixing with ordinary photons,
$\epsilon F'_{\mu\nu}F_{\mu\nu}$~\cite{Holdom:1985ag}.  The apparent
number 
count of CMB radiation can be modified by photon/dark
photon 
oscillation:
\begin{equation}
\frac{dn_{A}}{d\omega} \to \frac{dn_{A}}{d\omega} \times P_{A\to A} + \frac{dn_{A'}}{d\omega} \times P_{A'\to A}~,
\label{modified}
\end{equation}
where $P_{A\to A} = 1 - P_{A\to A'}$ is the photon survival
probability, while $ P_{A'\to A}$ is the probability of $A'\to A$
conversion.
Previous constraints from ${A \to A'}$ were
derived~\cite{Mirizzi:2009iz,Kunze:2015noa} using COBE-FIRAS
data~\cite{Fixsen:1996nj}.
The point of the present letter is that the RJ tail of the CMB can get
a significant boost due to the second term in (\ref{modified}) without
contradicting FIRAS\@.
 The physics of the 21\,cm line provides a useful tool to probe DR through the apparent modification of the low-energy tail of the CMB\@.  Previous studies~\cite{Nelson:2011sf,Arias:2012az,Graham:2015rva} have considered $A'\to A$ when $A'$ is cold DM instead of DR.

The EDGES experiment recently presented a tentative detection of 21\,cm
absorption coming from the interval of redshifts $z=15-20$~\cite{Bowman:2018yin}. The strength of the absorption signal is expected to be proportional to $1-T_{\rm CMB}/T_{s}$~\cite{Zaldarriaga:2003du}, where $T_{\rm CMB}$ counts the number of CMB photons interacting with the two-level hydrogen hyperfine system, and $T_s$ is the spin temperature.  The relevant photons have energy $\omega_0 = 5.9~\mu\mathrm{eV}$ at redshift $z \approx 17$, and therefore reside deep within the RJ tail,  $x \approx 1.4  \times 10^{-3}$. This is much lower energy than direct measurements from FIRAS, at $x > 1.2$~\cite{Fixsen:1996nj}, and ARCADE 2, which probes as low as $x = 0.056$~\cite{Fixsen:2009xn}.  Earlier measurements constrain $x \sim 0.02 - 0.04$ with larger uncertainties~\cite{1995ApL&C..32....3S,1995ApL&C..32....7B}.

The locations of the left/right boundaries of the claimed EDGES signal
agree with standard cosmological expectations, but the amount of
absorption seems to indicate a {\em more negative }
$1-T_{\rm CMB}/T_{s}$ temperature contrast than expected. Given that
the spin temperature $T_s$ cannot drop below the baryon temperature
$T_b$, a naive interpretation of this result could consist in
lower-than-expected $T_b$, or higher $T_{\rm CMB}$. Together with
related prior work~\cite{Tashiro:2014tsa,Munoz:2015bca}, a number of
possible models were
suggested~\cite{Barkana:2018lgd,Ewall-Wice:2018bzf,Barkana:2018qrx,Fialkov:2018xre,Fraser:2018acy},
which typically have difficulty passing other
constraints~\cite{Dvorkin:2013cea,Gluscevic:2017ywp,Xu:2018efh,Munoz:2018pzp,Berlin:2018sjs,DAmico:2018sxd}. Our
mechanism, the oscillation of non-thermal DR into visible photons (illustrated in Fig.~\ref{fig:Schematic}), can
easily accommodate EDGES consistent with other constraints. In the
remainder of this letter we provide more details on the suggested
mechanism, and identify the region of parameter space where 21\,cm
physics can provide the most sensitive probe of DR\@.

%\section{}

%%%%%%%%%%%%%%%%%%%%%%%%%%%%%%%%%%%%%%%%%%%%%%%%%%

\paragraph*{ Decay of unstable relics into dark radiation:}

The framework described in the introduction allows for significant flexibility with respect to the 
actual source of non-thermal soft DR\@. To give a  concrete realization of the proposed mechanism, 
we specify a model of unstable scalar particles, $a$, that couple to dark photons via an 
effective dimension five operator,
\begin{eqnarray}
\label{phiA'}
{\cal L}= \frac{1}{2}(\partial_\mu a)^2 - \frac{m_a^2}{2}a^2 + \frac{a}{4f_a} F'_{\mu\nu} \tilde F^{'\mu\nu}+{\cal L}_{AA'}~,
\end{eqnarray} 
where
$\tilde F'_{\mu\nu} = \frac{1}{2}
\epsilon^{\mu\nu\rho\sigma}F_{\rho\sigma}$, and the last term
describes the photon-dark photon Lagrangian with corresponding mass
and mixing terms for $A'$:
\begin{equation}
\label{AA'}
{\cal L}_{AA'}= -\frac{1}{4}F_{\mu\nu}^2 -\frac{1}{4}(F'_{\mu\nu})^2 -\frac{\epsilon}{2}F_{\mu\nu}F'_{\mu\nu}+\frac{1}{2}m_{A'}^2(A'_{\mu})^2~.
\end{equation} 
The cosmology of $a$ is model-dependent. To keep our discussion
general, we assume that an initial relic abundance of $a$ is present but we do not specify the method of non-thermal production, which could be misalignment or
  decay of cosmic strings.

The decay rate of  $a\to 2A'$ into two (transverse) dark photons $A'$ is given by,
\begin{equation}
\label{lifetime}
\Gamma_a =  \frac{m_a^3}{64\pi f_a^2} = \frac{3\times 10^{-4}}{\tau_{\rm U}} \left(\frac{m_a}{10^{-4}\,{\rm eV}}\right)^3
\left(\frac{100\,{\rm GeV}}{f_a}\right)^2\!\!.
\end{equation}
The lifetime, $\tau_a = 1/\Gamma_a$, can be either much longer or much shorter than the present age of the Universe, $\tau_{\rm U} \approx 13.8\times10^9~\mathrm{y}$.
In the case of short lifetimes, $\tau_a \ll \tau_{\rm U}$, the mass of $a$ should be chosen such that  the energy of the resulting $A'$ matches the CMB energy in the RJ tail at the time of decay.  We leave detailed study of this possibility for future work, and instead focus on the case of a cosmologically long-lived particle, $\tau_a \gg \tau_{\rm U}$, which is attractive because $a$ can naturally serve as DM\@.
If the mass of $a$ falls in the range $10^{-5}\,{\rm eV}<m_a< 10^{-1}\,{\rm eV}$,   
its decay can create significant modifications to the RJ tail of the CMB spectrum via $A' \rightarrow A$ oscillations.
It is worth noting that this overlaps with the mass range often invoked for axion DM.

 \begin{figure*}[!t!]
\centering
\includegraphics[width=0.38\textwidth]{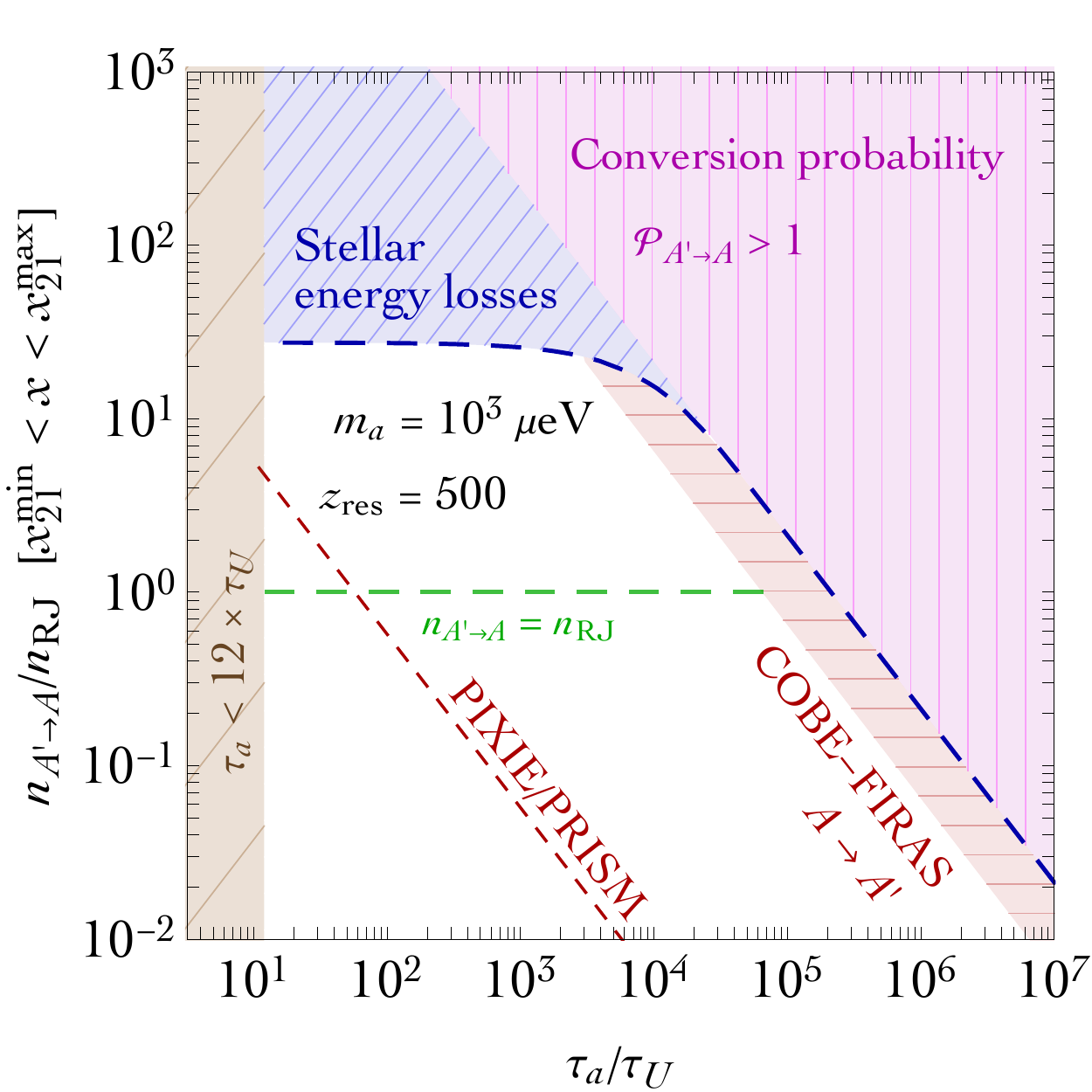} \hspace{0.65cm}%
 \raisebox{0.0cm}{\includegraphics[width=0.3775\textwidth]{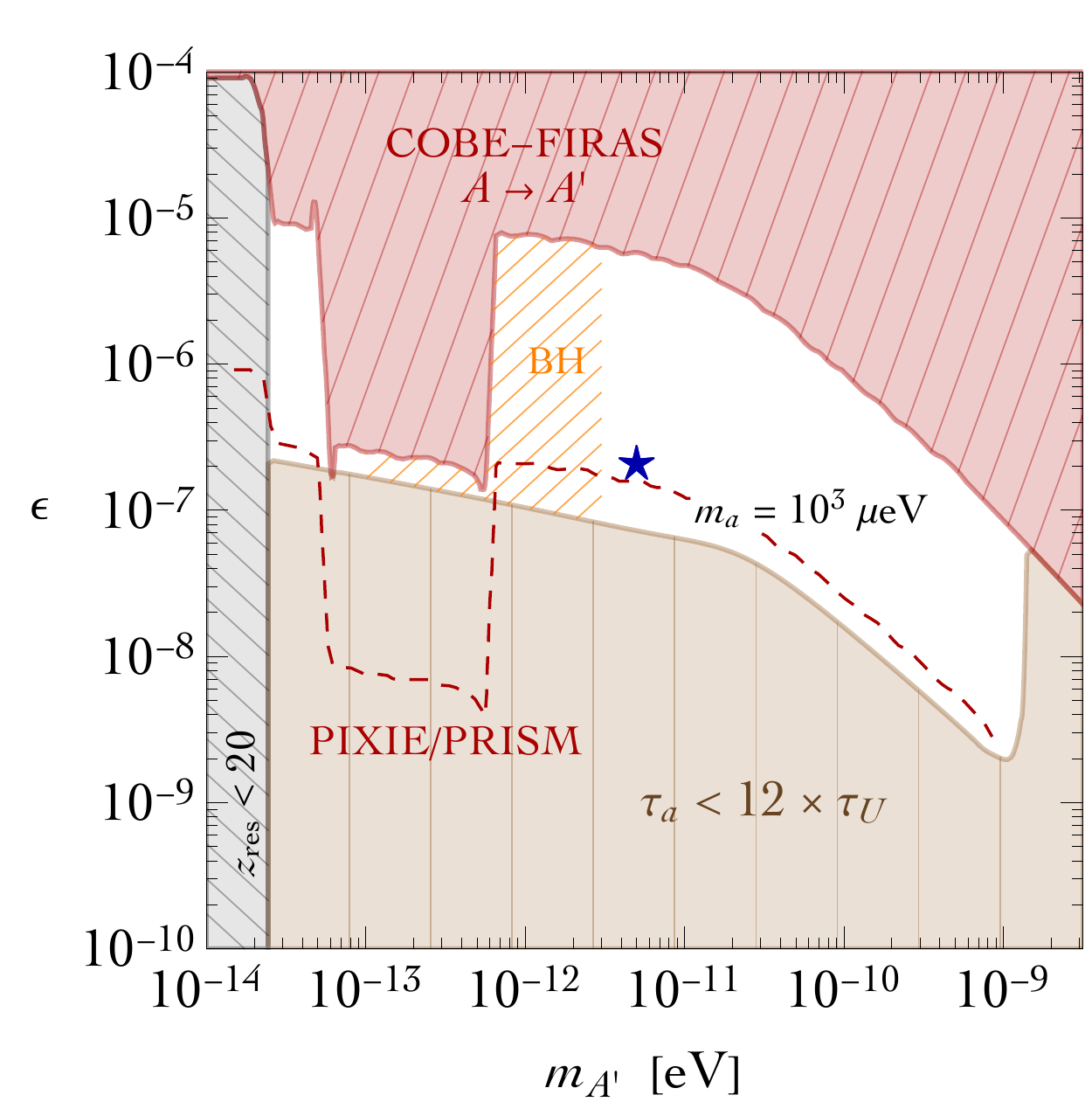}}%
\vspace{-.2 cm}
\caption{\label{fig:CoolingAndSpectrumDarkPhoton}
  {\em Left.}  Allowed region in DM lifetime versus number ratio of
  converted-to-original photons, $n_{A'\to A}/n_{\rm RJ}$, in the
  energy window $x_{21}^{\rm min} < x < x_{21}^{\rm max}$.  The blue
  region is excluded by stellar energy loss; the purple region would
  require $P_{A'\to A}>1$.  {\em Right.}  Allowed parameter space in
  the $(m_{A'}, \epsilon)$ plane; $\tau_{a}$ is chosen at each point
  such that $n_{A'\to A}/n_{\rm RJ} = 1$ in the 21\,cm signal
  region. In both plots $m_a=10^3 \mu\rm{eV}$, the brown region is
  excluded by a limit on the DM lifetime~\cite{Poulin:2016nat}, and the
  red region (red dashed line) is excluded by FIRAS (would be probed
  by PIXIE/PRISM) due to $A \rightarrow A'$
  oscillations~\cite{Kunze:2015noa}.  }
\end{figure*}

\paragraph*{ $A'\leftrightarrow A$ oscillation and constraints:}

All constraints on parameters of (\ref{phiA'}) and (\ref{AA'}) can be
divided into two groups: those that decouple as $m_{A'}\to0$, and
those that persist in the limit of a massless dark photon. The stellar
energy loss constraint due to $A'a$ pair production is in this second
category, as the in-medium transverse modes of photons can decay via
$A_T^*\to A'a$ even in the $m_{A'}\to0$ limit.  The approximate
emission rate is
$ Q_{A^*\to A'a} = \epsilon^2m_{A}^4 n_T / (96\pi f_a^2)$,
where $n_T$ is the number density of transverse plasmons (photons) and
$m_A$ is the plasma frequency, $m_A^2 = 4\pi \alpha n_e /m_e$.  We
recast the stellar bound on emission of Dirac neutrinos due to their
magnetic moment~\cite{Bernstein:1963qh, Raffelt:1992pi,Haft:1993jt} to
obtain the following bound:
$\epsilon\times f_a^{-1}< 2\times 10^{-9}\times {\rm GeV}^{-1}$.  The process $e^+ e^- \rightarrow a A'$  can lead to thermal populations of $a$ and $A'$ in the early Universe.  However the stellar constraint implies that this reaction drops out of equilibrium at temperatures above the electroweak scale, and therefore does not lead to an additional constraint from $N_{\rm eff}$.

The $\epsilon$-parameter is also constrained via $A\to A'$
oscillations \cite{Essig:2013lka}, and the resulting bound depends sensitively on
$m_{A'}$. Stellar energy losses via these oscillations are important
only for the higher mass range, $m_{A'} > 10^{-5}$\,eV, as the
emission is suppressed by $m_{A'}^2/m_{A}^2 \ll 1$ inside
stars~\cite{An:2013yfc,Redondo:2013lna}. Cosmological
$A\leftrightarrow A'$ oscillations may be significant {\em if} the
resonant condition is met, $m_{A'}=m_{A}(z)$, where $m_A(z)$ is the
plasma mass of photons at redshift
$z$~\cite{Mirizzi:2009iz,Kunze:2015noa}. In the course of cosmological
evolution
$m_{A}(z) \simeq 1.7 \times 10^{-14}\eV \times (1+z)^{3/2}
X^{1/2}_e(z) $ scans many orders of magnitude; $X_e$ is the free
electron fraction that we take from~\cite{Kunze:2015noa}.  For any
$m_{A'}$ in the range $10^{-14} - 10^{-9}$~eV, the resonance happens
at some redshift, $z_{\rm res}$ 
corresponding to cosmic time $t_{\rm res}$; see the left panel of
Fig.~\ref{fig:MassAndProb}. The resonance ensures that the probability
of oscillation is much larger than the vacuum value of
$\epsilon^2$. Following~\cite{Kuo:1989qe,Mirizzi:2009iz},
\begin{equation}
  \label{eq:conversionprobability}
P_{A\to A'}= P_{A'\to A}= \frac{\pi \epsilon^2 m_{A'}^2}{\omega}\times \left| \frac{d\log m_{A}^2}{dt}\right|_{t=t_{\rm res}}^{-1}.
\end{equation}
This expression is valid  in the limit
$P_{A' \rightarrow A} \ll 1$; 
when the probability saturates, we use the full expression of~\cite{Parke:1986jy}.
The probability of oscillation for RJ photons with $x \sim 10^{-3}$ can
be $10^3$ times larger than for typical CMB photons with $x \sim 1$,
because of the $\omega^{-1}$ scaling.  The redshift dependence of
(\ref{eq:conversionprobability}) is shown in the right panel of
Fig.~\ref{fig:MassAndProb}, assuming $x_0 = 1.4 \times 10^{-3}$.
 We multiply the conversion probability times the photon survival probability 
 to include the effect of free-free (bremsstrahlung) absorption. The latter is given by $P_s(x,z) \approx e^{-\tau_{\rm ff}(x,z)}$, where $\tau_{\rm ff}$ is the free-free absorption optical depth~\cite{Chluba:2015hma}. 
For $x\approx x_0$, photon absorption becomes relevant at redshift $z \gtrsim z_{\rm abs} = 1700$.
 
Before proceeding, we comment on another possible signature from
conversion at low redshifts, as the decay of $a$ inside clusters of
galaxies will lead to a flux of $A'$.  The latter could be converted
to regular photons as dark photons travel outside such a cluster, from
high to low density. The typical density of electrons inside clusters,
$n_e\sim 10^{-3}{\rm cm}^{-3}$~\cite{Bahcall:1995tf}, implies resonant
conversions when $m_{A'} \sim 10^{-12}$\,eV\@.
Thus, radio and microwave emission from clusters may contain new line
components; this will be addressed separately.

\paragraph*{ Dark age resonance and EDGES signal:}
For $z \leq z_{\rm abs}$, the Universe becomes transparent to
photons that are converted into the RJ tail of the CMB,
$x \sim 10^{-3}$, whereas for $z>z_{\rm abs}$ these soft photons are
efficiently absorbed~\cite{Chluba:2015hma}. Therefore, only dark
photons with $m_A < m_{A'}(z_{\rm abs}) \simeq 10^{-9}\,\eV$
will yield excess
radiation at 21\,cm. Focusing on a mono-chromatic injection of $A'$ with
cosmologically long lifetime $\tau_a > \tau_U$, the energy spectrum at
redshift $z$ reads,
\begin{align}
  \label{eq:DPspec}
  \frac{dn_{A'}}{d\omega} (\omega, z) =  \frac{2 \Omega_a \rho_c (1+z)^3  }{m_a \tau_a \omega H(\alpha-1)} \Theta(\alpha - 1 - z)~.
\end{align}
Here, $\rho_c $ is the critical density, $\Omega_a h^2 =  0.12$, and
 the Hubble rate, $H(z)$, is evaluated at redshift $\alpha-1$, where
$\alpha \equiv m_a (1+z)/(2\omega)$. 
The total number of injected $A'$
grows with cosmic time $t(z)$, and can eventually outgrow
$n_{\rm CMB}$ by a large margin,
$n_{A'}(z) = (6\,\eV/m_a)(t(z)/\tau_a) \times n_{\rm CMB}(z)$.  

Once the resonance condition is met at $z=z_{\rm res}$, a fraction of
$A'$ will be converted as per Eq.~(\ref{modified}). The final spectrum of converted photons at $z_{\rm res}$ is given by Eq.~(\ref{eq:DPspec}) multiplied by the conversion probability of Eq.~(\ref{eq:conversionprobability}).
To describe the spectrum at $z < z_{\rm res}$, it is convenient to switch to redshift-independent $x = \omega / T_{\rm CMB}$, noting that $dn / dx $ scales as $(1+z)^3$ to account for the expansion of the Universe.
The left panel of Fig.~\ref{fig:BoundMassAndSpectrumAfterConversion} shows the spectrum before and after conversion, in comparison to the RJ tail of the CMB, assuming $m_{a} = 10^{-3}$~eV, 
$\tau_{a} = 100\times \tau_{\rm U}$, $z_{\rm res} = 500$, and $\epsilon = 2.1\times 10^{-7}$.  For these parameters,  the total energy density of dark photons relative to the energy density of ordinary CMB photons at resonance is $6 \times 10^{-6}$, and a fraction $4 \times 10^{-4}$ of the dark photons oscillate into ordinary photons.

In order to identify models that can be tested with 21\,cm observations,  we compare the number density of converted photons, $n_{A' \rightarrow A}$, to the RJ density of the CMB, $n_{\rm RJ}$, within a relevant energy window.  We define this window to include all photons with a wavelength of  21\,cm within the redshift interval $z = 15-20$.  This is equivalent to requiring $x \in (x_{21}^{\rm min}, x_{21}^{\rm max}) = (1.2,1.6) \times 10^{-3}$.

The right panel of Fig.~\ref{fig:BoundMassAndSpectrumAfterConversion} shows the $\{ m_{A'},m_a \}$
parameter space relevant for the 21\,cm signal. Low values of $m_a$ give photons 
that are too soft to affect the 21\,cm absorption line, while high values produce photons at energies that (depending on 
other parameters) could be probed by FIRAS and ARCADE 2. Low and high limiting values for $m_{A'}$ originate from the requirement of
a resonance in the relevant redshift window, with some interval $10^{-13}-10^{-12}$\,eV possibly disfavored 
by black hole superradiance~\cite{Cardoso:2018tly} (see also~\cite{Pani:2012vp,Baryakhtar:2017ngi}). 
However, it is worth noting that these bounds were derived for a massive free dark photon without including interactions due to the mixing term, and their rigorous applicability in our case deserves a dedicated analysis.

We determine by how much the RJ photon
count can be increased due to $A'\to A$ conversion in the
$ 1.2\times 10^{-3}< x<1.6\times 10^{-3} $ interval.  The left panel of
Fig.~\ref{fig:CoolingAndSpectrumDarkPhoton} shows the allowed values
for the RJ photon increase, assuming $z_{\rm res} = 500$, when the lifetime $\tau_a$ of the DM particle
is varied.  We observe that the
photon count can be increased by more than an order of magnitude with
higher values limited by the stellar bound. If $n_{A'\to A}/n_{\rm RJ}$ is kept constant,
while $\tau_a$ is increased, the required value of $\epsilon$
is eventually excluded by CMB spectral distortions. Still, we
find that the unexpected strength of the EDGES signal, which would
require roughly $n_{A'\to A}/n_{\rm RJ}\simeq 1$, is easily met
over four orders of magnitude in the lifetime,
$\tau_a\sim (10-10^5)\tau_{\rm U}$. Finally, the right panel of the
same figure presents the $\{m_{A'},\epsilon\}$ parameter
space, where $\tau_a$ is chosen such that $n_{A'\to A}/n_{\rm RJ}$ is set to 1.
We find significant allowed parameter space, which can be mostly probed by the proposed PIXIE/PRISM experiments that are sensitive to spectral distortions~\cite{Kunze:2015noa}.

\paragraph*{Conclusions:}

We have shown that the RJ tail of the CMB spectrum can
be modified by light and weakly coupled New Physics particles/fields
without contradicting other cosmological or astrophysical constraints.
We have presented one such example where the resonant conversion of
non-thermal and numerous dark photons to ordinary photons leads to an
enhancement in the RJ tail of the CMB\@. The upcoming era of precision
21\,cm cosmology, as perhaps signaled by the first reported tentative
detection~\cite{Bowman:2018yin}, will open a window into such new
physics.

%%%%%%%%%%%%%%%%%%%%%%%%%%%%%%%%%%%%%%%%%%%%%%%%%%
%%%%%%%%%%%%%%%%%%%%%%%%%%%%%%%%%%%%%%%%%%%%%%%%%%
\paragraph*{Acknowledgments:}
We thank Yacine Ali-Ha\"imoud and Jens Chluba for helpful conversations.
MP and JTR acknowledge the financial support provided by CERN\@.
Research at Perimeter Institute is supported by the Government of
Canada through Industry Canada and by the Province of Ontario through
the Ministry of Economic Development \& Innovation. JP is supported by
the New Frontiers program of the Austrian Academy of Sciences.  JTR is supported by NSF CAREER grant PHY-1554858.

%%%%%%%%%%%%%%%%%%%%%%%%%%%%%%%%%%%%%%%%%%%%%%%%%%
%%%%%%%%%%%%%%%%%%%%%%%%%%%%%%%%%%%%%%%%%%%%%%%%%%

\bibliography{Dark21Refs}

%merlin.mbs apsrev4-1.bst 2010-07-25 4.21a (PWD, AO, DPC) hacked
%Control: key (0)
%Control: author (8) initials jnrlst
%Control: editor formatted (1) identically to author
%Control: production of article title (-1) disabled
%Control: page (0) single
%Control: year (1) truncated
%Control: production of eprint (0) enabled
\begin{thebibliography}{50}%
\makeatletter
\providecommand \@ifxundefined [1]{%
 \@ifx{#1\undefined}
}%
\providecommand \@ifnum [1]{%
 \ifnum #1\expandafter \@firstoftwo
 \else \expandafter \@secondoftwo
 \fi
}%
\providecommand \@ifx [1]{%
 \ifx #1\expandafter \@firstoftwo
 \else \expandafter \@secondoftwo
 \fi
}%
\providecommand \natexlab [1]{#1}%
\providecommand \enquote  [1]{``#1''}%
\providecommand \bibnamefont  [1]{#1}%
\providecommand \bibfnamefont [1]{#1}%
\providecommand \citenamefont [1]{#1}%
\providecommand \href@noop [0]{\@secondoftwo}%
\providecommand \href [0]{\begingroup \@sanitize@url \@href}%
\providecommand \@href[1]{\@@startlink{#1}\@@href}%
\providecommand \@@href[1]{\endgroup#1\@@endlink}%
\providecommand \@sanitize@url [0]{\catcode `\\12\catcode `\$12\catcode
  `\&12\catcode `\#12\catcode `\^12\catcode `\_12\catcode `\%12\relax}%
\providecommand \@@startlink[1]{}%
\providecommand \@@endlink[0]{}%
\providecommand \url  [0]{\begingroup\@sanitize@url \@url }%
\providecommand \@url [1]{\endgroup\@href {#1}{\urlprefix }}%
\providecommand \urlprefix  [0]{URL }%
\providecommand \Eprint [0]{\href }%
\providecommand \doibase [0]{http://dx.doi.org/}%
\providecommand \selectlanguage [0]{\@gobble}%
\providecommand \bibinfo  [0]{\@secondoftwo}%
\providecommand \bibfield  [0]{\@secondoftwo}%
\providecommand \translation [1]{[#1]}%
\providecommand \BibitemOpen [0]{}%
\providecommand \bibitemStop [0]{}%
\providecommand \bibitemNoStop [0]{.\EOS\space}%
\providecommand \EOS [0]{\spacefactor3000\relax}%
\providecommand \BibitemShut  [1]{\csname bibitem#1\endcsname}%
\let\auto@bib@innerbib\@empty
%</preamble>
\bibitem [{\citenamefont {Mather}\ \emph {et~al.}(1994)\citenamefont {Mather}
  \emph {et~al.}}]{Mather:1993ij}%
  \BibitemOpen
  \bibfield  {author} {\bibinfo {author} {\bibfnamefont {J.~C.}\ \bibnamefont
  {Mather}} \emph {et~al.},\ }\href {\doibase 10.1086/173574} {\bibfield
  {journal} {\bibinfo  {journal} {Astrophys. J.}\ }\textbf {\bibinfo {volume}
  {420}},\ \bibinfo {pages} {439} (\bibinfo {year} {1994})}\BibitemShut
  {NoStop}%
%%CITATION = ASJOA,420,439;%%
\bibitem [{\citenamefont {Hinshaw}\ \emph {et~al.}(2013)\citenamefont {Hinshaw}
  \emph {et~al.}}]{Hinshaw:2012aka}%
  \BibitemOpen
  \bibfield  {author} {\bibinfo {author} {\bibfnamefont {G.}~\bibnamefont
  {Hinshaw}} \emph {et~al.} (\bibinfo {collaboration} {WMAP}),\ }\href
  {\doibase 10.1088/0067-0049/208/2/19} {\bibfield  {journal} {\bibinfo
  {journal} {Astrophys. J. Suppl.}\ }\textbf {\bibinfo {volume} {208}},\
  \bibinfo {pages} {19} (\bibinfo {year} {2013})},\ \Eprint
  {http://arxiv.org/abs/1212.5226} {arXiv:1212.5226 [astro-ph.CO]} \BibitemShut
  {NoStop}%
%%CITATION = ARXIV:1212.5226;%%
\bibitem [{\citenamefont {Ade}\ \emph {et~al.}(2016)\citenamefont {Ade} \emph
  {et~al.}}]{Ade:2015xua}%
  \BibitemOpen
  \bibfield  {author} {\bibinfo {author} {\bibfnamefont {P.~A.~R.}\
  \bibnamefont {Ade}} \emph {et~al.} (\bibinfo {collaboration} {Planck}),\
  }\href {\doibase 10.1051/0004-6361/201525830} {\bibfield  {journal} {\bibinfo
   {journal} {Astron. Astrophys.}\ }\textbf {\bibinfo {volume} {594}},\
  \bibinfo {pages} {A13} (\bibinfo {year} {2016})},\ \Eprint
  {http://arxiv.org/abs/1502.01589} {arXiv:1502.01589 [astro-ph.CO]}
  \BibitemShut {NoStop}%
%%CITATION = ARXIV:1502.01589;%%
\bibitem [{\citenamefont {Abazajian}\ \emph {et~al.}(2016)\citenamefont
  {Abazajian} \emph {et~al.}}]{Abazajian:2016yjj}%
  \BibitemOpen
  \bibfield  {author} {\bibinfo {author} {\bibfnamefont {K.~N.}\ \bibnamefont
  {Abazajian}} \emph {et~al.} (\bibinfo {collaboration} {CMB-S4}),\ }\href@noop
  {} {\  (\bibinfo {year} {2016})},\ \Eprint {http://arxiv.org/abs/1610.02743}
  {arXiv:1610.02743 [astro-ph.CO]} \BibitemShut {NoStop}%
%%CITATION = ARXIV:1610.02743;%%
\bibitem [{\citenamefont {Furlanetto}\ \emph {et~al.}(2006)\citenamefont
  {Furlanetto}, \citenamefont {Oh},\ and\ \citenamefont
  {Briggs}}]{Furlanetto:2006jb}%
  \BibitemOpen
  \bibfield  {author} {\bibinfo {author} {\bibfnamefont {S.}~\bibnamefont
  {Furlanetto}}, \bibinfo {author} {\bibfnamefont {S.~P.}\ \bibnamefont {Oh}},
  \ and\ \bibinfo {author} {\bibfnamefont {F.}~\bibnamefont {Briggs}},\ }\href
  {\doibase 10.1016/j.physrep.2006.08.002} {\bibfield  {journal} {\bibinfo
  {journal} {Phys. Rept.}\ }\textbf {\bibinfo {volume} {433}},\ \bibinfo
  {pages} {181} (\bibinfo {year} {2006})},\ \Eprint
  {http://arxiv.org/abs/astro-ph/0608032} {arXiv:astro-ph/0608032 [astro-ph]}
  \BibitemShut {NoStop}%
%%CITATION = ASTRO-PH/0608032;%%
\bibitem [{\citenamefont {Jaeckel}\ and\ \citenamefont
  {Ringwald}(2010)}]{Jaeckel:2010ni}%
  \BibitemOpen
  \bibfield  {author} {\bibinfo {author} {\bibfnamefont {J.}~\bibnamefont
  {Jaeckel}}\ and\ \bibinfo {author} {\bibfnamefont {A.}~\bibnamefont
  {Ringwald}},\ }\href {\doibase 10.1146/annurev.nucl.012809.104433} {\bibfield
   {journal} {\bibinfo  {journal} {Ann. Rev. Nucl. Part. Sci.}\ }\textbf
  {\bibinfo {volume} {60}},\ \bibinfo {pages} {405} (\bibinfo {year} {2010})},\
  \Eprint {http://arxiv.org/abs/1002.0329} {arXiv:1002.0329 [hep-ph]}
  \BibitemShut {NoStop}%
%%CITATION = ARXIV:1002.0329;%%
\bibitem [{\citenamefont {Alexander}\ \emph {et~al.}(2016)\citenamefont
  {Alexander} \emph {et~al.}}]{Alexander:2016aln}%
  \BibitemOpen
  \bibfield  {author} {\bibinfo {author} {\bibfnamefont {J.}~\bibnamefont
  {Alexander}} \emph {et~al.}\ }(\bibinfo {year} {2016})\ \Eprint
  {http://arxiv.org/abs/1608.08632} {arXiv:1608.08632 [hep-ph]} \BibitemShut
  {NoStop}%
%%CITATION = ARXIV:1608.08632;%%
\bibitem [{\citenamefont {Battaglieri}\ \emph {et~al.}(2017)\citenamefont
  {Battaglieri} \emph {et~al.}}]{Battaglieri:2017aum}%
  \BibitemOpen
  \bibfield  {author} {\bibinfo {author} {\bibfnamefont {M.}~\bibnamefont
  {Battaglieri}} \emph {et~al.},\ }\href@noop {} {\  (\bibinfo {year}
  {2017})},\ \Eprint {http://arxiv.org/abs/1707.04591} {arXiv:1707.04591
  [hep-ph]} \BibitemShut {NoStop}%
%%CITATION = ARXIV:1707.04591;%%
\bibitem [{\citenamefont {Cherry}\ \emph {et~al.}(2015)\citenamefont {Cherry},
  \citenamefont {Frandsen},\ and\ \citenamefont {Shoemaker}}]{Cherry:2015oca}%
  \BibitemOpen
  \bibfield  {author} {\bibinfo {author} {\bibfnamefont {J.~F.}\ \bibnamefont
  {Cherry}}, \bibinfo {author} {\bibfnamefont {M.~T.}\ \bibnamefont
  {Frandsen}}, \ and\ \bibinfo {author} {\bibfnamefont {I.~M.}\ \bibnamefont
  {Shoemaker}},\ }\href {\doibase 10.1103/PhysRevLett.114.231303} {\bibfield
  {journal} {\bibinfo  {journal} {Phys. Rev. Lett.}\ }\textbf {\bibinfo
  {volume} {114}},\ \bibinfo {pages} {231303} (\bibinfo {year} {2015})},\
  \Eprint {http://arxiv.org/abs/1501.03166} {arXiv:1501.03166 [hep-ph]}
  \BibitemShut {NoStop}%
%%CITATION = ARXIV:1501.03166;%%
\bibitem [{\citenamefont {Cui}\ \emph {et~al.}(2017)\citenamefont {Cui},
  \citenamefont {Pospelov},\ and\ \citenamefont {Pradler}}]{Cui:2017ytb}%
  \BibitemOpen
  \bibfield  {author} {\bibinfo {author} {\bibfnamefont {Y.}~\bibnamefont
  {Cui}}, \bibinfo {author} {\bibfnamefont {M.}~\bibnamefont {Pospelov}}, \
  and\ \bibinfo {author} {\bibfnamefont {J.}~\bibnamefont {Pradler}},\
  }\href@noop {} {\  (\bibinfo {year} {2017})},\ \Eprint
  {http://arxiv.org/abs/1711.04531} {arXiv:1711.04531 [hep-ph]} \BibitemShut
  {NoStop}%
%%CITATION = ARXIV:1711.04531;%%
\bibitem [{\citenamefont {Berezhiani}\ \emph {et~al.}(2015)\citenamefont
  {Berezhiani}, \citenamefont {Dolgov},\ and\ \citenamefont
  {Tkachev}}]{Berezhiani:2015yta}%
  \BibitemOpen
  \bibfield  {author} {\bibinfo {author} {\bibfnamefont {Z.}~\bibnamefont
  {Berezhiani}}, \bibinfo {author} {\bibfnamefont {A.~D.}\ \bibnamefont
  {Dolgov}}, \ and\ \bibinfo {author} {\bibfnamefont {I.~I.}\ \bibnamefont
  {Tkachev}},\ }\href {\doibase 10.1103/PhysRevD.92.061303} {\bibfield
  {journal} {\bibinfo  {journal} {Phys. Rev.}\ }\textbf {\bibinfo {volume}
  {D92}},\ \bibinfo {pages} {061303} (\bibinfo {year} {2015})},\ \Eprint
  {http://arxiv.org/abs/1505.03644} {arXiv:1505.03644 [astro-ph.CO]}
  \BibitemShut {NoStop}%
%%CITATION = ARXIV:1505.03644;%%
\bibitem [{\citenamefont {Poulin}\ \emph {et~al.}(2016)\citenamefont {Poulin},
  \citenamefont {Serpico},\ and\ \citenamefont {Lesgourgues}}]{Poulin:2016nat}%
  \BibitemOpen
  \bibfield  {author} {\bibinfo {author} {\bibfnamefont {V.}~\bibnamefont
  {Poulin}}, \bibinfo {author} {\bibfnamefont {P.~D.}\ \bibnamefont {Serpico}},
  \ and\ \bibinfo {author} {\bibfnamefont {J.}~\bibnamefont {Lesgourgues}},\
  }\href {\doibase 10.1088/1475-7516/2016/08/036} {\bibfield  {journal}
  {\bibinfo  {journal} {JCAP}\ }\textbf {\bibinfo {volume} {1608}},\ \bibinfo
  {pages} {036} (\bibinfo {year} {2016})},\ \Eprint
  {http://arxiv.org/abs/1606.02073} {arXiv:1606.02073 [astro-ph.CO]}
  \BibitemShut {NoStop}%
%%CITATION = ARXIV:1606.02073;%%
\bibitem [{\citenamefont {Cardoso}\ \emph {et~al.}(2018)\citenamefont
  {Cardoso}, \citenamefont {Dias}, \citenamefont {Hartnett}, \citenamefont
  {Middleton}, \citenamefont {Pani},\ and\ \citenamefont
  {Santos}}]{Cardoso:2018tly}%
  \BibitemOpen
  \bibfield  {author} {\bibinfo {author} {\bibfnamefont {V.}~\bibnamefont
  {Cardoso}}, \bibinfo {author} {\bibfnamefont {O.~J.~C.}\ \bibnamefont
  {Dias}}, \bibinfo {author} {\bibfnamefont {G.~S.}\ \bibnamefont {Hartnett}},
  \bibinfo {author} {\bibfnamefont {M.}~\bibnamefont {Middleton}}, \bibinfo
  {author} {\bibfnamefont {P.}~\bibnamefont {Pani}}, \ and\ \bibinfo {author}
  {\bibfnamefont {J.~E.}\ \bibnamefont {Santos}},\ }\href@noop {} {\  (\bibinfo
  {year} {2018})},\ \Eprint {http://arxiv.org/abs/1801.01420} {arXiv:1801.01420
  [gr-qc]} \BibitemShut {NoStop}%
%%CITATION = ARXIV:1801.01420;%%
\bibitem [{\citenamefont {Holdom}(1986)}]{Holdom:1985ag}%
  \BibitemOpen
  \bibfield  {author} {\bibinfo {author} {\bibfnamefont {B.}~\bibnamefont
  {Holdom}},\ }\href {\doibase 10.1016/0370-2693(86)91377-8} {\bibfield
  {journal} {\bibinfo  {journal} {Phys. Lett.}\ }\textbf {\bibinfo {volume}
  {B166}},\ \bibinfo {pages} {196} (\bibinfo {year} {1986})}\BibitemShut
  {NoStop}%
%%CITATION = PHLTA,B166,196;%%
\bibitem [{\citenamefont {Mirizzi}\ \emph {et~al.}(2009)\citenamefont
  {Mirizzi}, \citenamefont {Redondo},\ and\ \citenamefont
  {Sigl}}]{Mirizzi:2009iz}%
  \BibitemOpen
  \bibfield  {author} {\bibinfo {author} {\bibfnamefont {A.}~\bibnamefont
  {Mirizzi}}, \bibinfo {author} {\bibfnamefont {J.}~\bibnamefont {Redondo}}, \
  and\ \bibinfo {author} {\bibfnamefont {G.}~\bibnamefont {Sigl}},\ }\href
  {\doibase 10.1088/1475-7516/2009/03/026} {\bibfield  {journal} {\bibinfo
  {journal} {JCAP}\ }\textbf {\bibinfo {volume} {0903}},\ \bibinfo {pages}
  {026} (\bibinfo {year} {2009})},\ \Eprint {http://arxiv.org/abs/0901.0014}
  {arXiv:0901.0014 [hep-ph]} \BibitemShut {NoStop}%
%%CITATION = ARXIV:0901.0014;%%
\bibitem [{\citenamefont {Kunze}\ and\ \citenamefont
  {Vázquez-Mozo}(2015)}]{Kunze:2015noa}%
  \BibitemOpen
  \bibfield  {author} {\bibinfo {author} {\bibfnamefont {K.~E.}\ \bibnamefont
  {Kunze}}\ and\ \bibinfo {author} {\bibfnamefont {M.~Ã.}\ \bibnamefont
  {Vázquez-Mozo}},\ }\href {\doibase 10.1088/1475-7516/2015/12/028} {\bibfield
   {journal} {\bibinfo  {journal} {JCAP}\ }\textbf {\bibinfo {volume} {1512}},\
  \bibinfo {pages} {028} (\bibinfo {year} {2015})},\ \Eprint
  {http://arxiv.org/abs/1507.02614} {arXiv:1507.02614 [astro-ph.CO]}
  \BibitemShut {NoStop}%
%%CITATION = ARXIV:1507.02614;%%
\bibitem [{\citenamefont {Fixsen}\ \emph {et~al.}(1996)\citenamefont {Fixsen},
  \citenamefont {Cheng}, \citenamefont {Gales}, \citenamefont {Mather},
  \citenamefont {Shafer},\ and\ \citenamefont {Wright}}]{Fixsen:1996nj}%
  \BibitemOpen
  \bibfield  {author} {\bibinfo {author} {\bibfnamefont {D.~J.}\ \bibnamefont
  {Fixsen}}, \bibinfo {author} {\bibfnamefont {E.~S.}\ \bibnamefont {Cheng}},
  \bibinfo {author} {\bibfnamefont {J.~M.}\ \bibnamefont {Gales}}, \bibinfo
  {author} {\bibfnamefont {J.~C.}\ \bibnamefont {Mather}}, \bibinfo {author}
  {\bibfnamefont {R.~A.}\ \bibnamefont {Shafer}}, \ and\ \bibinfo {author}
  {\bibfnamefont {E.~L.}\ \bibnamefont {Wright}},\ }\href {\doibase
  10.1086/178173} {\bibfield  {journal} {\bibinfo  {journal} {Astrophys. J.}\
  }\textbf {\bibinfo {volume} {473}},\ \bibinfo {pages} {576} (\bibinfo {year}
  {1996})},\ \Eprint {http://arxiv.org/abs/astro-ph/9605054}
  {arXiv:astro-ph/9605054 [astro-ph]} \BibitemShut {NoStop}%
%%CITATION = ASTRO-PH/9605054;%%
\bibitem [{\citenamefont {Nelson}\ and\ \citenamefont
  {Scholtz}(2011)}]{Nelson:2011sf}%
  \BibitemOpen
  \bibfield  {author} {\bibinfo {author} {\bibfnamefont {A.~E.}\ \bibnamefont
  {Nelson}}\ and\ \bibinfo {author} {\bibfnamefont {J.}~\bibnamefont
  {Scholtz}},\ }\href {\doibase 10.1103/PhysRevD.84.103501} {\bibfield
  {journal} {\bibinfo  {journal} {Phys. Rev.}\ }\textbf {\bibinfo {volume}
  {D84}},\ \bibinfo {pages} {103501} (\bibinfo {year} {2011})},\ \Eprint
  {http://arxiv.org/abs/1105.2812} {arXiv:1105.2812 [hep-ph]} \BibitemShut
  {NoStop}%
%%CITATION = ARXIV:1105.2812;%%
\bibitem [{\citenamefont {Arias}\ \emph {et~al.}(2012)\citenamefont {Arias},
  \citenamefont {Cadamuro}, \citenamefont {Goodsell}, \citenamefont {Jaeckel},
  \citenamefont {Redondo},\ and\ \citenamefont {Ringwald}}]{Arias:2012az}%
  \BibitemOpen
  \bibfield  {author} {\bibinfo {author} {\bibfnamefont {P.}~\bibnamefont
  {Arias}}, \bibinfo {author} {\bibfnamefont {D.}~\bibnamefont {Cadamuro}},
  \bibinfo {author} {\bibfnamefont {M.}~\bibnamefont {Goodsell}}, \bibinfo
  {author} {\bibfnamefont {J.}~\bibnamefont {Jaeckel}}, \bibinfo {author}
  {\bibfnamefont {J.}~\bibnamefont {Redondo}}, \ and\ \bibinfo {author}
  {\bibfnamefont {A.}~\bibnamefont {Ringwald}},\ }\href {\doibase
  10.1088/1475-7516/2012/06/013} {\bibfield  {journal} {\bibinfo  {journal}
  {JCAP}\ }\textbf {\bibinfo {volume} {1206}},\ \bibinfo {pages} {013}
  (\bibinfo {year} {2012})},\ \Eprint {http://arxiv.org/abs/1201.5902}
  {arXiv:1201.5902 [hep-ph]} \BibitemShut {NoStop}%
%%CITATION = ARXIV:1201.5902;%%
\bibitem [{\citenamefont {Graham}\ \emph {et~al.}(2016)\citenamefont {Graham},
  \citenamefont {Mardon},\ and\ \citenamefont {Rajendran}}]{Graham:2015rva}%
  \BibitemOpen
  \bibfield  {author} {\bibinfo {author} {\bibfnamefont {P.~W.}\ \bibnamefont
  {Graham}}, \bibinfo {author} {\bibfnamefont {J.}~\bibnamefont {Mardon}}, \
  and\ \bibinfo {author} {\bibfnamefont {S.}~\bibnamefont {Rajendran}},\ }\href
  {\doibase 10.1103/PhysRevD.93.103520} {\bibfield  {journal} {\bibinfo
  {journal} {Phys. Rev.}\ }\textbf {\bibinfo {volume} {D93}},\ \bibinfo {pages}
  {103520} (\bibinfo {year} {2016})},\ \Eprint
  {http://arxiv.org/abs/1504.02102} {arXiv:1504.02102 [hep-ph]} \BibitemShut
  {NoStop}%
%%CITATION = ARXIV:1504.02102;%%
\bibitem [{\citenamefont {Bowman}\ \emph {et~al.}(2018)\citenamefont {Bowman},
  \citenamefont {Rogers}, \citenamefont {Monsalve}, \citenamefont {Mozdzen},\
  and\ \citenamefont {Mahesh}}]{Bowman:2018yin}%
  \BibitemOpen
  \bibfield  {author} {\bibinfo {author} {\bibfnamefont {J.~D.}\ \bibnamefont
  {Bowman}}, \bibinfo {author} {\bibfnamefont {A.~E.~E.}\ \bibnamefont
  {Rogers}}, \bibinfo {author} {\bibfnamefont {R.~A.}\ \bibnamefont
  {Monsalve}}, \bibinfo {author} {\bibfnamefont {T.~J.}\ \bibnamefont
  {Mozdzen}}, \ and\ \bibinfo {author} {\bibfnamefont {N.}~\bibnamefont
  {Mahesh}},\ }\href {\doibase 10.1038/nature25792} {\bibfield  {journal}
  {\bibinfo  {journal} {Nature}\ }\textbf {\bibinfo {volume} {555}},\ \bibinfo
  {pages} {67} (\bibinfo {year} {2018})}\BibitemShut {NoStop}%
%%CITATION = NATUA,555,67;%%
\bibitem [{\citenamefont {Zaldarriaga}\ \emph {et~al.}(2004)\citenamefont
  {Zaldarriaga}, \citenamefont {Furlanetto},\ and\ \citenamefont
  {Hernquist}}]{Zaldarriaga:2003du}%
  \BibitemOpen
  \bibfield  {author} {\bibinfo {author} {\bibfnamefont {M.}~\bibnamefont
  {Zaldarriaga}}, \bibinfo {author} {\bibfnamefont {S.~R.}\ \bibnamefont
  {Furlanetto}}, \ and\ \bibinfo {author} {\bibfnamefont {L.}~\bibnamefont
  {Hernquist}},\ }\href {\doibase 10.1086/386327} {\bibfield  {journal}
  {\bibinfo  {journal} {Astrophys. J.}\ }\textbf {\bibinfo {volume} {608}},\
  \bibinfo {pages} {622} (\bibinfo {year} {2004})},\ \Eprint
  {http://arxiv.org/abs/astro-ph/0311514} {arXiv:astro-ph/0311514 [astro-ph]}
  \BibitemShut {NoStop}%
%%CITATION = ASTRO-PH/0311514;%%
\bibitem [{\citenamefont {Fixsen}\ \emph {et~al.}(2009)\citenamefont {Fixsen}
  \emph {et~al.}}]{Fixsen:2009xn}%
  \BibitemOpen
  \bibfield  {author} {\bibinfo {author} {\bibfnamefont {D.~J.}\ \bibnamefont
  {Fixsen}} \emph {et~al.},\ }\href {\doibase 10.1088/0004-637X/734/1/5} {\
  (\bibinfo {year} {2009}),\ 10.1088/0004-637X/734/1/5},\ \Eprint
  {http://arxiv.org/abs/0901.0555} {arXiv:0901.0555 [astro-ph.CO]} \BibitemShut
  {NoStop}%
%%CITATION = ARXIV:0901.0555;%%
\bibitem [{\citenamefont {{Staggs}}\ \emph {et~al.}(1995)\citenamefont
  {{Staggs}}, \citenamefont {{Jarosik}}, \citenamefont {{Wilkinson}},\ and\
  \citenamefont {{Wollack}}}]{1995ApL&C..32....3S}%
  \BibitemOpen
  \bibfield  {author} {\bibinfo {author} {\bibfnamefont {S.~T.}\ \bibnamefont
  {{Staggs}}}, \bibinfo {author} {\bibfnamefont {N.~C.}\ \bibnamefont
  {{Jarosik}}}, \bibinfo {author} {\bibfnamefont {D.~T.}\ \bibnamefont
  {{Wilkinson}}}, \ and\ \bibinfo {author} {\bibfnamefont {E.~J.}\ \bibnamefont
  {{Wollack}}},\ }\href@noop {} {\bibfield  {journal} {\bibinfo  {journal}
  {Astrophysical Letters and Communications}\ }\textbf {\bibinfo {volume}
  {32}},\ \bibinfo {pages} {3} (\bibinfo {year} {1995})}\BibitemShut {NoStop}%
\bibitem [{\citenamefont {{Bersanelli}}\ \emph {et~al.}(1995)\citenamefont
  {{Bersanelli}}, \citenamefont {{Smoot}}, \citenamefont {{Bensadoun}},
  \citenamefont {{de Amici}}, \citenamefont {{Limon}},\ and\ \citenamefont
  {{Levin}}}]{1995ApL&C..32....7B}%
  \BibitemOpen
  \bibfield  {author} {\bibinfo {author} {\bibfnamefont {M.}~\bibnamefont
  {{Bersanelli}}}, \bibinfo {author} {\bibfnamefont {G.~F.}\ \bibnamefont
  {{Smoot}}}, \bibinfo {author} {\bibfnamefont {M.}~\bibnamefont
  {{Bensadoun}}}, \bibinfo {author} {\bibfnamefont {G.}~\bibnamefont {{de
  Amici}}}, \bibinfo {author} {\bibfnamefont {M.}~\bibnamefont {{Limon}}}, \
  and\ \bibinfo {author} {\bibfnamefont {S.}~\bibnamefont {{Levin}}},\
  }\href@noop {} {\bibfield  {journal} {\bibinfo  {journal} {Astrophysical
  Letters and Communications}\ }\textbf {\bibinfo {volume} {32}},\ \bibinfo
  {pages} {7} (\bibinfo {year} {1995})}\BibitemShut {NoStop}%
\bibitem [{\citenamefont {Tashiro}\ \emph {et~al.}(2014)\citenamefont
  {Tashiro}, \citenamefont {Kadota},\ and\ \citenamefont
  {Silk}}]{Tashiro:2014tsa}%
  \BibitemOpen
  \bibfield  {author} {\bibinfo {author} {\bibfnamefont {H.}~\bibnamefont
  {Tashiro}}, \bibinfo {author} {\bibfnamefont {K.}~\bibnamefont {Kadota}}, \
  and\ \bibinfo {author} {\bibfnamefont {J.}~\bibnamefont {Silk}},\ }\href
  {\doibase 10.1103/PhysRevD.90.083522} {\bibfield  {journal} {\bibinfo
  {journal} {Phys. Rev.}\ }\textbf {\bibinfo {volume} {D90}},\ \bibinfo {pages}
  {083522} (\bibinfo {year} {2014})},\ \Eprint {http://arxiv.org/abs/1408.2571}
  {arXiv:1408.2571 [astro-ph.CO]} \BibitemShut {NoStop}%
%%CITATION = ARXIV:1408.2571;%%
\bibitem [{\citenamefont {Mu\~noz}\ \emph {et~al.}(2015)\citenamefont
  {Mu\~noz}, \citenamefont {Kovetz},\ and\ \citenamefont
  {Ali-Haïmoud}}]{Munoz:2015bca}%
  \BibitemOpen
  \bibfield  {author} {\bibinfo {author} {\bibfnamefont {J.~B.}\ \bibnamefont
  {Mu\~noz}}, \bibinfo {author} {\bibfnamefont {E.~D.}\ \bibnamefont {Kovetz}},
  \ and\ \bibinfo {author} {\bibfnamefont {Y.}~\bibnamefont {Ali-Haïmoud}},\
  }\href {\doibase 10.1103/PhysRevD.92.083528} {\bibfield  {journal} {\bibinfo
  {journal} {Phys. Rev.}\ }\textbf {\bibinfo {volume} {D92}},\ \bibinfo {pages}
  {083528} (\bibinfo {year} {2015})},\ \Eprint
  {http://arxiv.org/abs/1509.00029} {arXiv:1509.00029 [astro-ph.CO]}
  \BibitemShut {NoStop}%
%%CITATION = ARXIV:1509.00029;%%
\bibitem [{\citenamefont {Barkana}(2018)}]{Barkana:2018lgd}%
  \BibitemOpen
  \bibfield  {author} {\bibinfo {author} {\bibfnamefont {R.}~\bibnamefont
  {Barkana}},\ }\href {\doibase 10.1038/nature25791} {\bibfield  {journal}
  {\bibinfo  {journal} {Nature}\ }\textbf {\bibinfo {volume} {555}},\ \bibinfo
  {pages} {71} (\bibinfo {year} {2018})}\BibitemShut {NoStop}%
%%CITATION = NATUA,555,71;%%
\bibitem [{\citenamefont {Ewall-Wice}\ \emph {et~al.}(2018)\citenamefont
  {Ewall-Wice}, \citenamefont {Chang}, \citenamefont {Lazio}, \citenamefont
  {Dore}, \citenamefont {Seiffert},\ and\ \citenamefont
  {Monsalve}}]{Ewall-Wice:2018bzf}%
  \BibitemOpen
  \bibfield  {author} {\bibinfo {author} {\bibfnamefont {A.}~\bibnamefont
  {Ewall-Wice}}, \bibinfo {author} {\bibfnamefont {T.~C.}\ \bibnamefont
  {Chang}}, \bibinfo {author} {\bibfnamefont {J.}~\bibnamefont {Lazio}},
  \bibinfo {author} {\bibfnamefont {O.}~\bibnamefont {Dore}}, \bibinfo {author}
  {\bibfnamefont {M.}~\bibnamefont {Seiffert}}, \ and\ \bibinfo {author}
  {\bibfnamefont {R.~A.}\ \bibnamefont {Monsalve}},\ }\href@noop {} {\
  (\bibinfo {year} {2018})},\ \Eprint {http://arxiv.org/abs/1803.01815}
  {arXiv:1803.01815 [astro-ph.CO]} \BibitemShut {NoStop}%
%%CITATION = ARXIV:1803.01815;%%
\bibitem [{\citenamefont {Barkana}\ \emph {et~al.}(2018)\citenamefont
  {Barkana}, \citenamefont {Outmezguine}, \citenamefont {Redigolo},\ and\
  \citenamefont {Volansky}}]{Barkana:2018qrx}%
  \BibitemOpen
  \bibfield  {author} {\bibinfo {author} {\bibfnamefont {R.}~\bibnamefont
  {Barkana}}, \bibinfo {author} {\bibfnamefont {N.~J.}\ \bibnamefont
  {Outmezguine}}, \bibinfo {author} {\bibfnamefont {D.}~\bibnamefont
  {Redigolo}}, \ and\ \bibinfo {author} {\bibfnamefont {T.}~\bibnamefont
  {Volansky}},\ }\href@noop {} {\  (\bibinfo {year} {2018})},\ \Eprint
  {http://arxiv.org/abs/1803.03091} {arXiv:1803.03091 [hep-ph]} \BibitemShut
  {NoStop}%
%%CITATION = ARXIV:1803.03091;%%
\bibitem [{\citenamefont {Fialkov}\ \emph {et~al.}(2018)\citenamefont
  {Fialkov}, \citenamefont {Barkana},\ and\ \citenamefont
  {Cohen}}]{Fialkov:2018xre}%
  \BibitemOpen
  \bibfield  {author} {\bibinfo {author} {\bibfnamefont {A.}~\bibnamefont
  {Fialkov}}, \bibinfo {author} {\bibfnamefont {R.}~\bibnamefont {Barkana}}, \
  and\ \bibinfo {author} {\bibfnamefont {A.}~\bibnamefont {Cohen}},\
  }\href@noop {} {\  (\bibinfo {year} {2018})},\ \Eprint
  {http://arxiv.org/abs/1802.10577} {arXiv:1802.10577 [astro-ph.CO]}
  \BibitemShut {NoStop}%
%%CITATION = ARXIV:1802.10577;%%
\bibitem [{\citenamefont {Fraser}\ \emph {et~al.}(2018)\citenamefont {Fraser}
  \emph {et~al.}}]{Fraser:2018acy}%
  \BibitemOpen
  \bibfield  {author} {\bibinfo {author} {\bibfnamefont {S.}~\bibnamefont
  {Fraser}} \emph {et~al.},\ }\href@noop {} {\  (\bibinfo {year} {2018})},\
  \Eprint {http://arxiv.org/abs/1803.03245} {arXiv:1803.03245 [hep-ph]}
  \BibitemShut {NoStop}%
%%CITATION = ARXIV:1803.03245;%%
\bibitem [{\citenamefont {Dvorkin}\ \emph {et~al.}(2014)\citenamefont
  {Dvorkin}, \citenamefont {Blum},\ and\ \citenamefont
  {Kamionkowski}}]{Dvorkin:2013cea}%
  \BibitemOpen
  \bibfield  {author} {\bibinfo {author} {\bibfnamefont {C.}~\bibnamefont
  {Dvorkin}}, \bibinfo {author} {\bibfnamefont {K.}~\bibnamefont {Blum}}, \
  and\ \bibinfo {author} {\bibfnamefont {M.}~\bibnamefont {Kamionkowski}},\
  }\href {\doibase 10.1103/PhysRevD.89.023519} {\bibfield  {journal} {\bibinfo
  {journal} {Phys. Rev.}\ }\textbf {\bibinfo {volume} {D89}},\ \bibinfo {pages}
  {023519} (\bibinfo {year} {2014})},\ \Eprint {http://arxiv.org/abs/1311.2937}
  {arXiv:1311.2937 [astro-ph.CO]} \BibitemShut {NoStop}%
%%CITATION = ARXIV:1311.2937;%%
\bibitem [{\citenamefont {Gluscevic}\ and\ \citenamefont
  {Boddy}(2017)}]{Gluscevic:2017ywp}%
  \BibitemOpen
  \bibfield  {author} {\bibinfo {author} {\bibfnamefont {V.}~\bibnamefont
  {Gluscevic}}\ and\ \bibinfo {author} {\bibfnamefont {K.~K.}\ \bibnamefont
  {Boddy}},\ }\href@noop {} {\  (\bibinfo {year} {2017})},\ \Eprint
  {http://arxiv.org/abs/1712.07133} {arXiv:1712.07133 [astro-ph.CO]}
  \BibitemShut {NoStop}%
%%CITATION = ARXIV:1712.07133;%%
\bibitem [{\citenamefont {Xu}\ \emph {et~al.}(2018)\citenamefont {Xu},
  \citenamefont {Dvorkin},\ and\ \citenamefont {Chael}}]{Xu:2018efh}%
  \BibitemOpen
  \bibfield  {author} {\bibinfo {author} {\bibfnamefont {W.~L.}\ \bibnamefont
  {Xu}}, \bibinfo {author} {\bibfnamefont {C.}~\bibnamefont {Dvorkin}}, \ and\
  \bibinfo {author} {\bibfnamefont {A.}~\bibnamefont {Chael}},\ }\href@noop {}
  {\  (\bibinfo {year} {2018})},\ \Eprint {http://arxiv.org/abs/1802.06788}
  {arXiv:1802.06788 [astro-ph.CO]} \BibitemShut {NoStop}%
%%CITATION = ARXIV:1802.06788;%%
\bibitem [{\citenamefont {Mu\~noz}\ and\ \citenamefont
  {Loeb}(2018)}]{Munoz:2018pzp}%
  \BibitemOpen
  \bibfield  {author} {\bibinfo {author} {\bibfnamefont {J.~B.}\ \bibnamefont
  {Mu\~noz}}\ and\ \bibinfo {author} {\bibfnamefont {A.}~\bibnamefont {Loeb}},\
  }\href@noop {} {\  (\bibinfo {year} {2018})},\ \Eprint
  {http://arxiv.org/abs/1802.10094} {arXiv:1802.10094 [astro-ph.CO]}
  \BibitemShut {NoStop}%
%%CITATION = ARXIV:1802.10094;%%
\bibitem [{\citenamefont {Berlin}\ \emph {et~al.}(2018)\citenamefont {Berlin},
  \citenamefont {Hooper}, \citenamefont {Krnjaic},\ and\ \citenamefont
  {McDermott}}]{Berlin:2018sjs}%
  \BibitemOpen
  \bibfield  {author} {\bibinfo {author} {\bibfnamefont {A.}~\bibnamefont
  {Berlin}}, \bibinfo {author} {\bibfnamefont {D.}~\bibnamefont {Hooper}},
  \bibinfo {author} {\bibfnamefont {G.}~\bibnamefont {Krnjaic}}, \ and\
  \bibinfo {author} {\bibfnamefont {S.~D.}\ \bibnamefont {McDermott}},\
  }\href@noop {} {\  (\bibinfo {year} {2018})},\ \Eprint
  {http://arxiv.org/abs/1803.02804} {arXiv:1803.02804 [hep-ph]} \BibitemShut
  {NoStop}%
%%CITATION = ARXIV:1803.02804;%%
\bibitem [{\citenamefont {D'Amico}\ \emph {et~al.}(2018)\citenamefont
  {D'Amico}, \citenamefont {Panci},\ and\ \citenamefont
  {Strumia}}]{DAmico:2018sxd}%
  \BibitemOpen
  \bibfield  {author} {\bibinfo {author} {\bibfnamefont {G.}~\bibnamefont
  {D'Amico}}, \bibinfo {author} {\bibfnamefont {P.}~\bibnamefont {Panci}}, \
  and\ \bibinfo {author} {\bibfnamefont {A.}~\bibnamefont {Strumia}},\
  }\href@noop {} {\  (\bibinfo {year} {2018})},\ \Eprint
  {http://arxiv.org/abs/1803.03629} {arXiv:1803.03629 [astro-ph.CO]}
  \BibitemShut {NoStop}%
%%CITATION = ARXIV:1803.03629;%%
\bibitem [{\citenamefont {Bernstein}\ \emph {et~al.}(1963)\citenamefont
  {Bernstein}, \citenamefont {Ruderman},\ and\ \citenamefont
  {Feinberg}}]{Bernstein:1963qh}%
  \BibitemOpen
  \bibfield  {author} {\bibinfo {author} {\bibfnamefont {J.}~\bibnamefont
  {Bernstein}}, \bibinfo {author} {\bibfnamefont {M.}~\bibnamefont {Ruderman}},
  \ and\ \bibinfo {author} {\bibfnamefont {G.}~\bibnamefont {Feinberg}},\
  }\href {\doibase 10.1103/PhysRev.132.1227} {\bibfield  {journal} {\bibinfo
  {journal} {Phys. Rev.}\ }\textbf {\bibinfo {volume} {132}},\ \bibinfo {pages}
  {1227} (\bibinfo {year} {1963})}\BibitemShut {NoStop}%
%%CITATION = PHRVA,132,1227;%%
\bibitem [{\citenamefont {Raffelt}\ and\ \citenamefont
  {Weiss}(1992)}]{Raffelt:1992pi}%
  \BibitemOpen
  \bibfield  {author} {\bibinfo {author} {\bibfnamefont {G.}~\bibnamefont
  {Raffelt}}\ and\ \bibinfo {author} {\bibfnamefont {A.}~\bibnamefont
  {Weiss}},\ }\href@noop {} {\bibfield  {journal} {\bibinfo  {journal} {Astron.
  Astrophys.}\ }\textbf {\bibinfo {volume} {264}},\ \bibinfo {pages} {536}
  (\bibinfo {year} {1992})}\BibitemShut {NoStop}%
%%CITATION = AAEJA,264,536;%%
\bibitem [{\citenamefont {Haft}\ \emph {et~al.}(1994)\citenamefont {Haft},
  \citenamefont {Raffelt},\ and\ \citenamefont {Weiss}}]{Haft:1993jt}%
  \BibitemOpen
  \bibfield  {author} {\bibinfo {author} {\bibfnamefont {M.}~\bibnamefont
  {Haft}}, \bibinfo {author} {\bibfnamefont {G.}~\bibnamefont {Raffelt}}, \
  and\ \bibinfo {author} {\bibfnamefont {A.}~\bibnamefont {Weiss}},\ }\href
  {\doibase 10.1086/173978} {\bibfield  {journal} {\bibinfo  {journal}
  {Astrophys. J.}\ }\textbf {\bibinfo {volume} {425}},\ \bibinfo {pages} {222}
  (\bibinfo {year} {1994})},\ \bibinfo {note} {[Erratum: Astrophys.
  J.438,1017(1995)]},\ \Eprint {http://arxiv.org/abs/astro-ph/9309014}
  {arXiv:astro-ph/9309014 [astro-ph]} \BibitemShut {NoStop}%
%%CITATION = ASTRO-PH/9309014;%%
\bibitem [{\citenamefont {Essig}\ \emph {et~al.}(2013)\citenamefont {Essig}
  \emph {et~al.}}]{Essig:2013lka}%
  \BibitemOpen
  \bibfield  {author} {\bibinfo {author} {\bibfnamefont {R.}~\bibnamefont
  {Essig}} \emph {et~al.},\ }in\ \href
  {http://inspirehep.net/record/1263039/files/arXiv:1311.0029.pdf} {\emph
  {\bibinfo {booktitle} {{Proceedings, 2013 Community Summer Study on the
  Future of U.S. Particle Physics: Snowmass on the Mississippi (CSS2013):
  Minneapolis, MN, USA, July 29-August 6, 2013}}}}\ (\bibinfo {year} {2013})\
  \Eprint {http://arxiv.org/abs/1311.0029} {arXiv:1311.0029 [hep-ph]}
  \BibitemShut {NoStop}%
%%CITATION = ARXIV:1311.0029;%%
\bibitem [{\citenamefont {An}\ \emph {et~al.}(2013)\citenamefont {An},
  \citenamefont {Pospelov},\ and\ \citenamefont {Pradler}}]{An:2013yfc}%
  \BibitemOpen
  \bibfield  {author} {\bibinfo {author} {\bibfnamefont {H.}~\bibnamefont
  {An}}, \bibinfo {author} {\bibfnamefont {M.}~\bibnamefont {Pospelov}}, \ and\
  \bibinfo {author} {\bibfnamefont {J.}~\bibnamefont {Pradler}},\ }\href
  {\doibase 10.1016/j.physletb.2013.07.008} {\bibfield  {journal} {\bibinfo
  {journal} {Phys. Lett.}\ }\textbf {\bibinfo {volume} {B725}},\ \bibinfo
  {pages} {190} (\bibinfo {year} {2013})},\ \Eprint
  {http://arxiv.org/abs/1302.3884} {arXiv:1302.3884 [hep-ph]} \BibitemShut
  {NoStop}%
%%CITATION = ARXIV:1302.3884;%%
\bibitem [{\citenamefont {Redondo}\ and\ \citenamefont
  {Raffelt}(2013)}]{Redondo:2013lna}%
  \BibitemOpen
  \bibfield  {author} {\bibinfo {author} {\bibfnamefont {J.}~\bibnamefont
  {Redondo}}\ and\ \bibinfo {author} {\bibfnamefont {G.}~\bibnamefont
  {Raffelt}},\ }\href {\doibase 10.1088/1475-7516/2013/08/034} {\bibfield
  {journal} {\bibinfo  {journal} {JCAP}\ }\textbf {\bibinfo {volume} {1308}},\
  \bibinfo {pages} {034} (\bibinfo {year} {2013})},\ \Eprint
  {http://arxiv.org/abs/1305.2920} {arXiv:1305.2920 [hep-ph]} \BibitemShut
  {NoStop}%
%%CITATION = ARXIV:1305.2920;%%
\bibitem [{\citenamefont {Kuo}\ and\ \citenamefont
  {Pantaleone}(1989)}]{Kuo:1989qe}%
  \BibitemOpen
  \bibfield  {author} {\bibinfo {author} {\bibfnamefont {T.-K.}\ \bibnamefont
  {Kuo}}\ and\ \bibinfo {author} {\bibfnamefont {J.~T.}\ \bibnamefont
  {Pantaleone}},\ }\href {\doibase 10.1103/RevModPhys.61.937} {\bibfield
  {journal} {\bibinfo  {journal} {Rev. Mod. Phys.}\ }\textbf {\bibinfo {volume}
  {61}},\ \bibinfo {pages} {937} (\bibinfo {year} {1989})}\BibitemShut
  {NoStop}%
%%CITATION = RMPHA,61,937;%%
\bibitem [{\citenamefont {Parke}(1986)}]{Parke:1986jy}%
  \BibitemOpen
  \bibfield  {author} {\bibinfo {author} {\bibfnamefont {S.~J.}\ \bibnamefont
  {Parke}},\ }\bibfield  {booktitle} {\emph {\bibinfo {booktitle}
  {{Proceedings, 23RD International Conference on High Energy Physics, JULY
  16-23, 1986, Berkeley, CA}}},\ }\href {\doibase 10.1103/PhysRevLett.57.1275}
  {\bibfield  {journal} {\bibinfo  {journal} {Phys. Rev. Lett.}\ }\textbf
  {\bibinfo {volume} {57}},\ \bibinfo {pages} {1275} (\bibinfo {year}
  {1986})}\BibitemShut {NoStop}%
%%CITATION = PRLTA,57,1275;%%
\bibitem [{\citenamefont {Chluba}(2015)}]{Chluba:2015hma}%
  \BibitemOpen
  \bibfield  {author} {\bibinfo {author} {\bibfnamefont {J.}~\bibnamefont
  {Chluba}},\ }\href {\doibase 10.1093/mnras/stv2243} {\bibfield  {journal}
  {\bibinfo  {journal} {Mon. Not. Roy. Astron. Soc.}\ }\textbf {\bibinfo
  {volume} {454}},\ \bibinfo {pages} {4182} (\bibinfo {year} {2015})},\ \Eprint
  {http://arxiv.org/abs/1506.06582} {arXiv:1506.06582 [astro-ph.CO]}
  \BibitemShut {NoStop}%
%%CITATION = ARXIV:1506.06582;%%
\bibitem [{\citenamefont {Bahcall}(1995)}]{Bahcall:1995tf}%
  \BibitemOpen
  \bibfield  {author} {\bibinfo {author} {\bibfnamefont {N.~A.}\ \bibnamefont
  {Bahcall}},\ }in\ \href@noop {} {\emph {\bibinfo {booktitle} {{13th Jerusalem
  Winter School in Theoretical Physics: Formation of Structure in the Universe
  Jerusalem, Israel, 27 December 1995 - 5 January 1996}}}}\ (\bibinfo {year}
  {1995})\ \Eprint {http://arxiv.org/abs/astro-ph/9611148}
  {arXiv:astro-ph/9611148 [astro-ph]} \BibitemShut {NoStop}%
%%CITATION = ASTRO-PH/9611148;%%
\bibitem [{\citenamefont {Pani}\ \emph {et~al.}(2012)\citenamefont {Pani},
  \citenamefont {Cardoso}, \citenamefont {Gualtieri}, \citenamefont {Berti},\
  and\ \citenamefont {Ishibashi}}]{Pani:2012vp}%
  \BibitemOpen
  \bibfield  {author} {\bibinfo {author} {\bibfnamefont {P.}~\bibnamefont
  {Pani}}, \bibinfo {author} {\bibfnamefont {V.}~\bibnamefont {Cardoso}},
  \bibinfo {author} {\bibfnamefont {L.}~\bibnamefont {Gualtieri}}, \bibinfo
  {author} {\bibfnamefont {E.}~\bibnamefont {Berti}}, \ and\ \bibinfo {author}
  {\bibfnamefont {A.}~\bibnamefont {Ishibashi}},\ }\href {\doibase
  10.1103/PhysRevLett.109.131102} {\bibfield  {journal} {\bibinfo  {journal}
  {Phys. Rev. Lett.}\ }\textbf {\bibinfo {volume} {109}},\ \bibinfo {pages}
  {131102} (\bibinfo {year} {2012})},\ \Eprint {http://arxiv.org/abs/1209.0465}
  {arXiv:1209.0465 [gr-qc]} \BibitemShut {NoStop}%
%%CITATION = ARXIV:1209.0465;%%
\bibitem [{\citenamefont {Baryakhtar}\ \emph {et~al.}(2017)\citenamefont
  {Baryakhtar}, \citenamefont {Lasenby},\ and\ \citenamefont
  {Teo}}]{Baryakhtar:2017ngi}%
  \BibitemOpen
  \bibfield  {author} {\bibinfo {author} {\bibfnamefont {M.}~\bibnamefont
  {Baryakhtar}}, \bibinfo {author} {\bibfnamefont {R.}~\bibnamefont {Lasenby}},
  \ and\ \bibinfo {author} {\bibfnamefont {M.}~\bibnamefont {Teo}},\ }\href
  {\doibase 10.1103/PhysRevD.96.035019} {\bibfield  {journal} {\bibinfo
  {journal} {Phys. Rev.}\ }\textbf {\bibinfo {volume} {D96}},\ \bibinfo {pages}
  {035019} (\bibinfo {year} {2017})},\ \Eprint
  {http://arxiv.org/abs/1704.05081} {arXiv:1704.05081 [hep-ph]} \BibitemShut
  {NoStop}%
%%CITATION = ARXIV:1704.05081;%%
\end{thebibliography}%

\end{document}